\documentclass[sigconf,10pt,nonacm,authordraf,screen]{acmart}

\AtBeginDocument{%
  \providecommand\BibTeX{{%
    \normalfont B\kern-0.5em{\scshape i\kern-0.25em b}\kern-0.8em\TeX}}}

\usepackage{amsmath,amsopn,amsthm,wasysym,mathtools}
\usepackage{float}
\usepackage{subcaption}
\usepackage{endnotes,microtype,xspace,graphicx,fancyvrb,multirow}
\usepackage{booktabs}
\usepackage{enumitem}
\usepackage[labelfont=bf,font=small,skip=5pt]{caption}
\usepackage[ruled,linesnumbered]{algorithm2e}
\SetKwInput{kwInit}{Init}
\SetKwComment{Comment}{\textnormal{/*} }{ \textnormal{*/}}

\usepackage{fp}
\usepackage{siunitx}

\usepackage[acronym,nohypertypes={acronym,notation}]{glossaries}

\usepackage{balance}

\usepackage{tikz}
\usetikzlibrary{calc}
\usetikzlibrary{positioning}
\usetikzlibrary{fit}
\usetikzlibrary{shapes}
\usetikzlibrary{trees}
\usetikzlibrary{arrows}
\usetikzlibrary{overlay-beamer-styles}
\usepackage{xcolor}

\usepackage{changepage}

\newacronym{DoS}{DoS}{denial-of-service}
\newcommand{\sys}{\mbox{\textsc{SeMalloc}}\xspace}
\newcommand{\sema}{\mbox{\texttt{SemaType}}\xspace}

\newcommand{\nid}{\mbox{\texttt{nID}}\xspace}
\newcommand{\rid}{\mbox{\texttt{rID}}\xspace}

\long\def\ignore#1{}

\newcommand{\glibc}{glibc\xspace}

\newcommand{\cc}[1]{\mbox{\smaller[0.5]\texttt{#1}}}

\input{code/fmt}

\def\Snospace~{\S{}}

\newif\ifdraft\drafttrue
\newif\ifnotes\notestrue
\ifdraft\else\notesfalse\fi

\input{glyphtounicode}
\newcommand{\squishlist}{
\begin{itemize}[noitemsep,nolistsep]
  \setlength{\itemsep}{-0pt}
}
\newcommand{\squishend}{
  \end{itemize}
}

\newcommand*\WC[1]{%
\begin{tikzpicture}[baseline=(C.base)]
\node[draw,circle,inner sep=0.2pt](C) {#1};
\end{tikzpicture}}

\usepackage{xstring}
\newcommand{\PP}[1]{
\vspace{2px}
\noindent{\bf \IfEndWith{#1}{.}{#1}{#1.}}
}

\newcommand{\boxbeg}{
\vspace{2px}
\noindent\begin{tabular}{|l|}\hline
\begin{minipage}{\linewidth - 2.5ex}
\vspace{2px}
\noindent
}

\newcommand{\boxend}{
\vspace{2px}
\end{minipage}\\ \hline
\end{tabular}
\vspace{3pt}
}

\newenvironment{myquote}[1]%
  {\list{}{\leftmargin=#1\rightmargin=#1}\item[]}%
  {\endlist}

\begin{document}

\title{\sys: Semantics-Informed Memory Allocator}

\ifdefined\DRAFT
 \pagestyle{fancyplain}
 \lhead{Rev.~\therev}
 \rhead{\thedate}
 \cfoot{\thepage\ of \pageref{LastPage}}
\fi



\author{Ruizhe Wang}
\email{ruizhe.wang@uwaterloo.ca}
\affiliation{%
\institution{University of Waterloo}
\country{Canada}
}

\author{Meng Xu}
\email{meng.xu.cs@uwaterloo.ca}
\affiliation{%
\institution{University of Waterloo}
  \country{Canada}
}

\author{N. Asokan}
\email{asokan@acm.org}
\affiliation{%
  \institution{University of Waterloo}
  \country{Canada}
}



\setcopyright{acmlicensed}
\copyrightyear{2018}
\acmYear{2018}
\acmDOI{XXXXXXX.XXXXXXX}

\acmConference[CCS '24]{the 2024 ACM SIGSAC Conference on Computer and Communications
Security}{October 14--18,
  2024}{Salt Lake City, UT}
\date{}
\begin{abstract}
Use-after-free (UAF)
is a critical and prevalent problem
in memory unsafe languages.
While many solutions have been proposed,
balancing
security, run-time cost, and memory overhead
(an impossible trinity) is hard.

In this paper,
we show one way to balance the trinity
by passing more semantics
about the heap object to the allocator for it
to make informed allocation decisions.
More specifically,
we propose a new notion of
thread-, context-, and flow-sensitive ``type'', \sema,
to capture the semantics and
prototype a \sema-based allocator that
aims for the best trade-off amongst the impossible trinity.
In \sys,
only heap objects allocated from the same call site and
via the same function call stack
can possibly share a virtual memory address,
which effectively stops type-confusion attacks and
makes UAF vulnerabilities harder to exploit.

Through extensive empirical evaluation,
we show that \sys is realistic:
(a)  \sys is effective in
thwarting all real-world vulnerabilities we tested;
(b) benchmark programs run even slightly faster with \sys
than the default heap allocator,
at a memory overhead averaged from 41\% to 84\%; and
(c) \sys balances security and overhead strictly better
than other closely related works.
\end{abstract}

\maketitle
\sloppy

\section{Introduction}
\label{s:intro}

Heap vulnerabilities are common
in memory unsafe languages like C and C++.
Exploiting these vulnerabilities,
attackers can inflict 
denial-of-service,
information leakage, or
arbitrary code execution.
Use-after-free (UAF)
is a typical class of heap vulnerabilities
that have received special attention due to both
its prevalence and
the number and variety of powerful exploits
it enables~\cite{sok-eternal-war}.

UAF happens when a memory chunk is accessed after it is freed.
More specifically,
freeing a heap object renders
all pointers to this object (or parts thereof)
\emph{dangling}.
Any memory access through a dangling pointer
can lead to \emph{undefined behavior}
according to the C standard~\cite{c23}.

There is a wealth of prior research
intended to address UAF vulnerabilities
(see \autoref{s:bg} for an exposition)
and pros and cons can be found in
each theme of UAF-mitigation techniques.
For example,
some allocators suffer from incomplete protection while
others may
incur prohibitively high run-time or memory overhead.
While no allocation strategy is unquestionably superior
in mitigating UAF vulnerabilities,
\emph{type-based} allocation,
which permits the reuse of memory chunks
only among allocations of the same type,
seems to be a promising direction
and is the focus of this paper.

Although
type-based allocation provides imperfect protection only,
the protection is more predictable
than entroy-based allocators and more importantly,
the protection can be achieved with reasonable overheads.
However,
existing type-based allocators are either coarse-grained in its definition of type~\cite{cling,typeaftertype} leading to weaker protection,
or extremely fine-grained,
treating each heap object as a different ``type"~\cite{ffmalloc}, and leading to complete protection at a very high cost.
Therefore,
a gap remains in the design space for type-based allocators
to balance between security and overheads.

The goal of this paper is to find a sweet spot
in the design space of type-based allocation
that achieves sufficiently high protection
without excessive overhead.
More specifically,
we present \sys,
a type-based UAF-mitigating allocator
that operates on a new definition of \textbf{type}
at its core:
\begin{myquote}{0.1in}
    Two heap objects are of the same type
    \emph{if and only if}
    they are
    (a) allocated from the same allocation site
    (e.g., a specific \cc{malloc} call), and
    (b) the allocation call is invoked
    under the same call stack,
    modulo recursion.
\end{myquote}

\noindent
To avoid confusion with the conventional notion
of type in programming languages,
we denote our "type" definition \sema.
For programs hardened with \sys,
UAF can only occur between heap objects
 of the same \sema.

\sys's run-time and memory overheads
are low enough to make it suitable
for real-world use.
For instance,
on SPEC CPU 2017,
\sys incurs an average run-time overhead of -0.6\%
which is faster than
MarkUs~\cite{markus},
MineSweeper~\cite{minesweep},
and DangZero~\cite{dangzero}
(by giving up protection against
UAF within the same \sema),
and is similar to TypeAfterType~\cite{typeaftertype}
(with improved security)
and PUMM~\cite{pumm}
(with improved usability).
\sys incurs an average memory overhead of 61.0\%
which is much lower than FFMalloc~\cite{ffmalloc}
(again, by giving up protection against
UAF within the same \sema)
but is higher than TypeAfterType
due to improved type sensitivity
(hence security).

\PP{Summary}
We claim the following contributions:

\begin{itemize}[noitemsep,topsep=0pt,leftmargin=*]
    \item A callout that the ``type''
    in type-based heap allocator can be defined differently
    and does not need to be a native type in the
    programming language
    (\autoref{ss:bg-type});
    
    \item The design and implementation of a new type-based memory allocator \sys which uses 
    \sema, a carefully designed ``type'', to
    target a sweet spot between
    sensitivity (which decides security) and
    performance (which is affected by tracking overhead) (\autoref{s:overview}-\autoref{s:design}); and
    
    %
   \item A thorough evaluation of \sys
   (with an \href{https://drive.google.com/file/d/1kLSihIANcnIEpBao64xf0yyIMEend0I0/view?usp=drive_link}{open artifact})
   showing that it
   successfully detects all real-world attacks
   we tested (\autoref{s:sec}) with marginal overheads (\autoref{s:eval}).
   %
\end{itemize}
\vspace{5pt}

\section{A Mini SoK on UAF}
\label{s:bg}

We present a mini
systematization of knowledge (SoK)
on techniques that exploit or mitigate
UAF vulnerabilities.
The purposes of this SoK is
to help position \sys in the research landscape, and
to identify
synergies among UAF-mitigating strategies and
defence-in-depth opportunities.

\subsection{Exploiting UAF Vulnerabilities}
\label{ss:bg-exploit}

UAF is generally considered as a \emph{temporal} memory error,
i.e., an error that occurs following
a specific temporal order of events.
In the context of UAF,
the events include
allocation (e.g., \cc{malloc}),
de-allocation (e.g., \cc{free}),
read, and write.
Fortunately (or unfortunately),
for most programs,
there are plenty of such events
in their original code logic;
all an attacker needs to do is
to find and trigger the correct sequence of events
to mount an attack
without code injection.
\autoref{fig:uaf-demo} is a
crafted example to show how
a dangling pointer can be exploited differently
with different event orderings.

Formally,
if a new object $N$
(accessible through a fresh pointer $p$)
is allocated over the heap location
previously occupied by a freed object $O$
(which leaves a dangling pointer $q$),
then one of the following cases can happen:

\begin{itemize}[noitemsep,leftmargin=\parindent]
    \item[A] a read through $p$ breaches the confidentiality of $O$,
    although this is usually called \emph{uninitialized read},
    which is generally not a concern of UAF
    and can be mitigated via zeroing allocations~\cite{lee:dangnull, dangzero}
    or other techniques~\cite{unisan,safeinit};

    \item[B] a write through $p$ breaches the integrity of $O$,
    as the written content can be subsequently read through $q$
    which can compromise the execution context where $q$ is used;
    %

    \item[C] a read through $q$ breaches the confidentiality of $N$
    which can be used to leak sensitive information such as
    pointer addresses (to break ASLR~\cite{aslr1, aslr2}) or secret data;
    %

    \item[D] a write through $q$ breaches the integrity of $N$,
    as the written content can be subsequently read through $p$
    which can compromise the execution context where $p$ is used;
    %

    \item[E] a free through $q$ de-allocates $N$ entirely,
    and yet, the heap allocator cannot block it
    if $p$ and $q$ are the same integer
    representing memory addresses
    (a free through $p$ is legit).
\end{itemize}

\noindent
Exploit B-E can all be found in~\autoref{fig:uaf-demo}.
Again,
note that
from an attacker's point of view,
exploiting a UAF bug does not
require code injection.
Instead, an attacker can craft a
``weird machine''~\cite{weird-machine}
by merely re-purposing operations
involving the inadvertent alias pair ($p$, $q$)
in the original code logic.
Intuitively,
the more operations an attacker can re-purpose,
the more useful a UAF can be in launching attacks.
In the extreme case where
any new object can be allocated
over the heap location accessible
by the dangling pointer $q$,
this UAF is effectively
an arbitrary read/write exploit primitive.

\begin{figure}[t]
    \centering
    \small
    \hfill\begin{minipage}[c]{.925\linewidth}\input{code/uaf.c}\end{minipage}
    \caption{A hypothetical example to illustrate UAF exploits. \\
    \textnormal{
        Exploit-B: line 16--5--7--17 $\quad\;\rightarrow$ arbitrary code execution \\
        Exploit-C: line 16--5--6--8--20 $\,\rightarrow$ information leak \\
        Exploit-D: line 16--5--19--9 $\quad\;\rightarrow$ arbitrary code execution \\
        Exploit-E: line 16--5--21 $\quad\quad\;\rightarrow$ $p$ is de-allocated and dangling
    }}
    \label{fig:uaf-demo}
\end{figure}

On a side note,
Exploit-E is different from
what is conventionally known as \emph{double free}
which arises when the old pointer $q$ is freed twice
without the allocation of a new heap object $N$.
Double free vulnerabilities can be mitigated  cheaply
by maintaining a set of freed and yet-to-be allocated memory addresses~\cite{guarder,slimguard,dieharder}
as a top-up of other UAF-mitigation strategies.

\PP{Type confusion}
The methodology to
exploit the UAF bug in~\autoref{fig:uaf-demo} is also known as
\emph{type confusion} or \emph{type manipulation},
which is arguably the most popular way to exploit a UAF bug,
especially when an object type involved contains a function pointer.
However,
type confusion is not the only way to exploit a UAF;
and more importantly, a UAF bug can be exploited
even when the two objects involved have the same type,
as shown in~\autoref{fig:uaf-demo-same-type}.
Despite the fact that
both the victim pointer $p$ and dangling pointer $q$
share the same type,
one can still leak sensitive data via $p$ or
cause \cc{\_\_real\_fn} to be called in \cc{register\_fake} and vice versa.

\PP{Multi-threading and race conditions}
Although~\autoref{fig:uaf-demo} and \ref{fig:uaf-demo-same-type}
are demonstrated in a multi-threaded setting,
and indeed many exploits in the wild
require some form of race condition to work~\cite{race-uaf},
multi-threading is not a strict requirement to exploit a UAF bug,
as long as the attackers can find a similar sequence of events in a sequential execution,
as showcased in~\cite{CVE-2015-6831, CVE-2015-6835, CVE-2018-11496, python-24613, mruby-4001, yasm-91}.

\begin{figure}[t]
    \centering
    \small
    \hfill\begin{minipage}[c]{.925\linewidth}\input{code/uaf-same-type.c}\end{minipage}
    \caption{A hypothetical example to illustrate UAF exploits
    against objects of the same type.}
    \label{fig:uaf-demo-same-type}
\end{figure}

\subsection{Mitigating UAF Vulnerabilities}
\label{ss:bg-defense}

Attackers' view on
how to exploit UAF vulnerabilities~(\autoref{ss:bg-exploit})
also sheds light on how to mitigate UAF,
which has been extensively researched.
In fact,
existing techniques line up
nicely as layered protection
against UAF vulnerabilities:

\textbf{A)
Invalidate dangling pointers},
which breaks the foundation of any UAF exploits.
In literature,
this has been achieved via
various creative techniques,
including:

\begin{itemize}[noitemsep,topsep=0pt,leftmargin=*]
    \item Track pointer derivation at runtime
        and nullify all associated pointers
        upon object de-allocation~\cite{lee:dangnull,dangsan,FreeSentry,heapexpo};

    \item Treat pointers as capabilities to access memory
    (instead of integers)
    and \cc{free} revokes the capability~\cite{cornucopia,dangzero,oscar}.
\end{itemize}

\noindent
Prior works in this category
have demonstrated complete protection against
heap-based UAF
but may pay the price of
compatibility (e.g., CHERI~\cite{cheri}),
kernel privilege~\cite{dangzero},
high overhead (e.g., 80\% run-time overhead for DangNull~\cite{lee:dangnull}), or
subtle complexities as shown in HeapExpo~\cite{heapexpo}.

\textbf{B1)
Prevent heap objects
from being allocated over any pointer that might be dangling},
which can be achieved by 
tracking pointer derivation~\cite{crcount} or
sweeping all stored pointers~\cite{minesweep,Efficientsweep,markus}.
In other words,
these allocators
do not trust the \cc{free} request from developers;
instead,
they de-allocate memory
only when ``absolutely'' safe.
Hence, allocators in this category
can achieve complete protection against heap-based UAF
(modulo subtle pointer propagation flows~\cite{heapexpo}).
However,
they arguably introduce high overheads.
For example,
MarkUs~\cite{markus}
more than doubles the run-time on the PARSEC benchmark
(see~\autoref{ss:eval_macro}).

\textbf{B2)
Prevent heap objects that an attacker targets}
(\emph{victim objects})
\textbf{from being allocated over a dangling pointer}.
This is essentially a weaker version of B1 and
entrusts allocators to decide
which class of objects should or should never be allocated
on a specific free memory chunk.
Intuitively,
the ideal allocator would never place
a victim object over a freed memory chunk with
attacker-controlled dangling pointers.

Unfortunately,
this ideal allocator cannot exist,
as there is no way for an allocator to tell
which object can be a \emph{victim}
(i.e., a valuable target for attackers)
among all allocated objects,
even with information from
static or dynamic program analysis.
Hence, in theory, perfect UAF-mitigation
is not possible with this approach.

However,
if an allocator knows enough about the \emph{semantics}
of allocated objects,
it can place objects of different semantics
into different and isolated pools.
In this way,
a dangling pointer of certain semantics controlled by the attacker
can only be used to access newly allocated objects
bearing the same semantics.
This is commonly known as
\emph{type-based allocation}
which makes UAF exploitation harder
by confining what attackers can do
after obtaining a dangling pointer.
Not surprisingly,
allocators in this category~\cite{cling,typeaftertype,ffmalloc},
including \sys,
differ on their definition of semantics or type,
which we give an in-depth reflection in~\autoref{ss:bg-type}.

PUMM~\cite{pumm} proposes that 
the completion of a ``task'' can be a clear signal 
to de-allocate freed memory accumulated in the ended task
such that the freed memory can be re-allocated
in a new \emph{task}
which is irrelevant to any previous tasks.
As tasks can have arbitrarily-defined boundary
(e.g., one iteration of a loop is one task),
PUMM effectively encodes temporal information into type and
in theory, can be complementary to type-based allocators~\cite{cling,typeaftertype}
including \sys.

\textbf{B3)
Confine allocations or
reduce the level of certainty on
when and which a victim object
will be allocated over a dangling pointer.}
Allocators in this category~\cite{dieharder,slimguard,guarder}
typically leverage on multiple randomization schemes
to boost entropy, such as
big bag of pages (BIBOP)-based~\cite{bibop} allocation and
delayed freelist.
Randomization provides
a probabilistic defense against UAF exploits
by lowering their success rates.
However,
entropy-based allocators can still be prone to
information leaks or heap fengshui~\cite{heap-feng-shui}
which reduces the level of entropy in practice.

With that said,
entropy-based allocators typically incur
low overhead (both in run-time and memory).
For example,
Guarder reports a 3\% run-time overhead and
27\% memory overhead
with its highest level of entropy configuration
while SlimGuard~\cite{slimguard}
further reduces the memory footprint.

\textbf{C)
Validate a pointer upon use.}
This line of work checks whether a pointer is
safe for read/write operation upon dereference and
can detect UAF attempts on the spot.
Achieving this typically requires
heavy instrumentation on instructions that
may access memory through a pointer
which significantly outnumber
\cc{malloc} and \cc{free} operations.
This explains the high overhead~\cite{valgrind,asan,giantsan}
even amongst the works designed to run in production~\cite{ptauth,pactight,vik,pacman,fatpointer}
(e.g., 20\% run-time overhead for Vik~\cite{vik}).

\PP{Summary}
Prior works in categories A, B1, and C
can mitigate all heap-based UAF attacks
(assuming perfect implementation)
but might also
incur excessive overheads or
require special hardware or kernel modification.
Type- or entropy-based secure allocators
(categories B2, B3) incur smaller overheads
at the expense of incomplete protection.

\emph{Defense-in-depth}:
While all building blocks for a layered
UAF defense has been proposed,
to the best of our knowledge,
there is no real-world allocator that combines them,
in all or in parts.
This SoK shows an opportunity
to provide a defense-in-depth solution
that holistically integrates memory allocators
of all themes of defences.

\subsection{A Reflection on Semantics And Type}
\label{ss:bg-type}

In programming languages,
\emph{``type''} is typically considered as
a token that encodes some ``semantics'' of an object.
As briefly discussed in~\autoref{ss:bg-defense},
a type-based heap allocator
confines the types of objects
a dangling pointer might ever access~\cite{cling,typeaftertype,ffmalloc}.
More specifically,
if a freed object is of type $\mathbb{T}$,
only objects of the same type $\mathbb{T}$
can then be allocated over the free chunk.
Intuitively,
type-based allocation provides
a \emph{tunable} defense
against UAF with a clear security and performance trade-off---all
by varying the definition of ``type''.

While there are many ways to define types
(hence the research on type systems~\cite{type-system}),
one particularly useful angle in the context of
type-based allocation is the \emph{sensitivity} of a type,
i.e., how well a type can distinguish heap allocations
occurring under different execution states.
The insight is that:
\textbf{objects allocated under the same or similar execution states
are expected to behave similarly in the program},
and such behaviors are essentially the semantics of the objects,
which serve as the ``type'' in type-based allocation.

Borrowing sensitivity notions
from program analysis,
we can define type sensitivity from the following perspectives:

\PP{Flow-sensitive}
If a function
is invoked in two places within the same function,
a flow-sensitive type will differentiate these two function calls.
To illustrate,
in the code below,
the two \cc{malloc} calls are different
under a flow-sensitive scheme.
\begin{adjustwidth}{.03\textwidth}{0cm}{\small \begin{Verbatim}[commandchars=\\\{\},numbers=left,firstnumber=1,stepnumber=1,codes={\catcode`\$=3\catcode`\^=7\catcode`\_=8\relax}]
\PY{k+kt}{void}\PY{+w}{ }\PY{n+nf}{foo}\PY{p}{(}\PY{p}{)}\PY{+w}{ }\PY{p}{\PYZob{}}
\PY{+w}{  }\PY{k+kt}{void}\PY{+w}{ }\PY{o}{*}\PY{n}{p}\PY{+w}{ }\PY{o}{=}\PY{+w}{ }\PY{n}{malloc}\PY{p}{(}\PY{k}{sizeof}\PY{p}{(}\PY{k+kt}{int}\PY{p}{)}\PY{p}{)}\PY{p}{;}
\PY{+w}{  }\PY{k+kt}{void}\PY{+w}{ }\PY{o}{*}\PY{n}{q}\PY{+w}{ }\PY{o}{=}\PY{+w}{ }\PY{n}{malloc}\PY{p}{(}\PY{k}{sizeof}\PY{p}{(}\PY{k+kt}{int}\PY{p}{)}\PY{p}{)}\PY{p}{;}
\PY{p}{\PYZcb{}}
\end{Verbatim}
}\end{adjustwidth}

\PP{Path-sensitive}
If a function
is reached via different control-flow paths
within a function
(modulo loop),
a path-sensitive type will differentiate
these two execution paths.
To illustrate,
in the code below,
the \cc{malloc} call might allocate
objects of different types depending
on the boolean \cc{cond}.
\begin{adjustwidth}{.03\textwidth}{0cm}{\small \begin{Verbatim}[commandchars=\\\{\},numbers=left,firstnumber=1,stepnumber=1,codes={\catcode`\$=3\catcode`\^=7\catcode`\_=8\relax}]
\PY{k+kt}{void}\PY{+w}{ }\PY{n+nf}{foo}\PY{p}{(}\PY{k+kt}{bool}\PY{+w}{ }\PY{n}{cond}\PY{p}{)}\PY{+w}{ }\PY{p}{\PYZob{}}
\PY{+w}{  }\PY{k+kt}{size\PYZus{}t}\PY{+w}{ }\PY{n}{len}\PY{+w}{ }\PY{o}{=}\PY{+w}{ }\PY{k}{sizeof}\PY{p}{(}\PY{k+kt}{int}\PY{p}{)}\PY{p}{;}
\PY{+w}{  }\PY{k}{if}\PY{+w}{ }\PY{p}{(}\PY{n}{cond}\PY{p}{)}\PY{+w}{ }\PY{p}{\PYZob{}}\PY{+w}{    }\PY{n}{len}\PY{+w}{ }\PY{o}{=}\PY{+w}{ }\PY{k}{sizeof}\PY{p}{(}\PY{k+kt}{long}\PY{p}{)}\PY{p}{;}  \PY{+w}{  }\PY{p}{\PYZcb{}}
\PY{+w}{  }\PY{k+kt}{void}\PY{+w}{ }\PY{o}{*}\PY{n}{p}\PY{+w}{ }\PY{o}{=}\PY{+w}{ }\PY{n}{malloc}\PY{p}{(}\PY{n}{len}\PY{p}{)}\PY{p}{;}
\PY{p}{\PYZcb{}}
\end{Verbatim}
}\end{adjustwidth}

\PP{Context-sensitive}
If a function
is reached via different call traces (modulo recursion),
a context-sensitive type
will differentiate these two calling contexts.
To illustrate,
in the code below,
the \cc{malloc} call under contexts
[\cc{foo} $\rightarrow$ \cc{wrapper}] and
[\cc{bar} $\rightarrow$ \cc{wrapper}]
allocate objects of different types.
\begin{adjustwidth}{.03\textwidth}{0cm}{\small \begin{Verbatim}[commandchars=\\\{\},numbers=left,firstnumber=1,stepnumber=1,codes={\catcode`\$=3\catcode`\^=7\catcode`\_=8\relax}]
\PY{k+kt}{void}\PY{+w}{ }\PY{n+nf}{wrapper}\PY{p}{(}\PY{k+kt}{size\PYZus{}t}\PY{+w}{ }\PY{n}{len}\PY{p}{)}\PY{+w}{ }\PY{p}{\PYZob{}}
\PY{+w}{  }\PY{k+kt}{void}\PY{+w}{ }\PY{o}{*}\PY{n}{p}\PY{+w}{ }\PY{o}{=}\PY{+w}{ }\PY{n}{malloc}\PY{p}{(}\PY{n}{len}\PY{p}{)}\PY{p}{;}
\PY{p}{\PYZcb{}}
\PY{k+kt}{void}\PY{+w}{ }\PY{n+nf}{foo}\PY{p}{(}\PY{p}{)}\PY{+w}{ }\PY{p}{\PYZob{}}\PY{+w}{  }\PY{n}{wrapper}\PY{p}{(}\PY{k}{sizeof}\PY{p}{(}\PY{k+kt}{int}\PY{p}{)}\PY{p}{)}\PY{p}{;}  \PY{p}{\PYZcb{}}
\PY{k+kt}{void}\PY{+w}{ }\PY{n+nf}{bar}\PY{p}{(}\PY{p}{)}\PY{+w}{ }\PY{p}{\PYZob{}}\PY{+w}{  }\PY{n}{wrapper}\PY{p}{(}\PY{k}{sizeof}\PY{p}{(}\PY{k+kt}{int}\PY{p}{)}\PY{p}{)}\PY{p}{;}  \PY{p}{\PYZcb{}}
\end{Verbatim}
}\end{adjustwidth}

\PP{Thread-sensitive}
If a function
is invoked in different threads,
a thread-sensitive type will differentiate the threads.
To illustrate,
in the code below,
the \cc{malloc} call under the two threads
allocates objects of different types.
\begin{adjustwidth}{.03\textwidth}{0cm}{\small \begin{Verbatim}[commandchars=\\\{\},numbers=left,firstnumber=1,stepnumber=1,codes={\catcode`\$=3\catcode`\^=7\catcode`\_=8\relax}]
\PY{k+kt}{void}\PY{+w}{ }\PY{o}{*}\PY{n+nf}{thread}\PY{p}{(}\PY{k+kt}{void}\PY{+w}{ }\PY{o}{*}\PY{n}{ptr}\PY{p}{)}\PY{+w}{ }\PY{p}{\PYZob{}}
\PY{+w}{  }\PY{k+kt}{void}\PY{+w}{ }\PY{o}{*}\PY{n}{p}\PY{+w}{ }\PY{o}{=}\PY{+w}{ }\PY{n}{malloc}\PY{p}{(}\PY{k}{sizeof}\PY{p}{(}\PY{k+kt}{int}\PY{p}{)}\PY{p}{)}\PY{p}{;}
\PY{p}{\PYZcb{}}
\PY{k+kt}{void}\PY{+w}{ }\PY{n+nf}{foo}\PY{p}{(}\PY{p}{)}\PY{+w}{ }\PY{p}{\PYZob{}}
\PY{+w}{  }\PY{n}{pthread\PYZus{}t}\PY{+w}{ }\PY{n}{t1}\PY{p}{,}\PY{+w}{ }\PY{n}{t2}\PY{p}{;}
\PY{+w}{  }\PY{n}{pthread\PYZus{}create}\PY{p}{(}\PY{o}{\PYZam{}}\PY{n}{t1}\PY{p}{,}\PY{+w}{ }\PY{n+nb}{NULL}\PY{p}{,}\PY{+w}{ }\PY{o}{*}\PY{k+kr}{thread}\PY{p}{,}\PY{+w}{ }\PY{n+nb}{NULL}\PY{p}{)}\PY{p}{;}
\PY{+w}{  }\PY{n}{pthread\PYZus{}create}\PY{p}{(}\PY{o}{\PYZam{}}\PY{n}{t2}\PY{p}{,}\PY{+w}{ }\PY{n+nb}{NULL}\PY{p}{,}\PY{+w}{ }\PY{o}{*}\PY{k+kr}{thread}\PY{p}{,}\PY{+w}{ }\PY{n+nb}{NULL}\PY{p}{)}\PY{p}{;}
\PY{p}{\PYZcb{}}
\end{Verbatim}
}\end{adjustwidth}

\PP{Sensitivity in cyclic control-flow structures}
In loops and recursive calls,
the sensitivity is typically classified as:
\begin{itemize}[noitemsep,topsep=0pt,leftmargin=*]
    \item \emph{Unbounded},
    where different iterations of a loop or recursion
    yield different types.
    
    \item \emph{Bounded},
    where different iterations of a loop or recursion
    yield different types, up to a pre-defined limit.
    
    \item \emph{Insensitive},
    where different iterations of a loop or recursion
    yield the same type.
\end{itemize}

\PP{Finding the right sensitivity level}
For a type-based heap allocator to be secure yet practical,
finding the right sensitivity level is the key.

A type definition with higher sensitivity implies
a smaller set of object types
a dangling pointer may point to.
In this regard,
the default \cc{glibc} allocator~\cite{glibc-malloc}
in most Linux-based systems
is (almost) completely insensitive.
Regarding the two closely related works,
the type definition in Cling~\cite{cling} 
adopts a weaker form of context-sensitivity,
where the context is defined as
two innermost return addresses
on the current call stack---an
approximation to call trace.
Type-after-Type~\cite{typeaftertype}
is based on statically-inferred unqualified types
native to the programming language
enriched by \cc{malloc} wrapper trace.

And yet,
for a type-based allocator,
it is not necessarily true that
more sensitivity is better.
To illustrate,
the type definition with the highest sensitivity is to
treat every heap object as a different type.
This effectively means that
a heap allocator will never reclaim memory---an
impractical approach,
as a long-running program may allocate and free
an endless number of heap objects yet
the virtual memory address space has a limit
(e.g., 48-bit on x86).
FFMalloc~\cite{ffmalloc} is a close approximation
to this extreme approach and
incurs a large memory overhead despite the fact that
it still reclaims virtual pages.
A path-, context-, and thread-sensitive type qualifier
will be extremely sensitive as well,
and yet,
tracking path sensitivity requires
instrumentation at basic block granularity,
which adds a significant overhead.

\PP{Conclusion}
While prior works have tried to
address UAF vulnerabilities,
the challenge of finding the right balance between
the level of protection and incurred overhead
remains.
Certainly manually porting the program has the potential of
achieving the trinity of optimizing all memory,
run-time, and security~\cite{fatpointer},
in the rest of this paper,
we present our approach
towards finding such a balance without this aid.
\section{Capture Semantics with \NoCaseChange{\sema}}
\label{s:overview}

We now introduce \sema,
a type qualifier~\cite{type-qualifier}
tailored to capture the semantics
of heap allocations,
and showcase how to deduce \sema 
at runtime through a concrete example.

\subsection{Defining \sema}
\label{ss:sema_define}

\sema is a
\emph{thread-, context- and flow-sensitive
type qualifier}
over the standard type system
of the underlying programming language
(e.g., LLVM IR in \sys)
with
\emph{bounded sensitivity for recursions} and
\emph{no sensitivity for loops}
(sensitivity levels defined in~\autoref{ss:bg-type}).

Informally,
in a more operative description,
two heap objects are of the same \sema
\emph{if and only if} they are:
\begin{itemize}[noitemsep,topsep=0pt,leftmargin=*]
    \item allocated from the same allocation site
    (e.g., the very same \cc{malloc} call
    in the source code); and
    
    \item allocated
        under the same call stack, modulo recursion.
\end{itemize}
In the presence of recursive calls,
\sema differentiates call traces inside
each strongly connected component
(SCC, representing a group of recursive calls)
in the call graph up to a fixed limit.
In \sys, this bound is $2^{14}$
different call traces overall (see~\autoref{fig:encoding}).

\noindent
\textbf{Deducing \sema.}$\;$
In theory,
the \sema of every heap allocation
can be deduced at compile-time
by inlining all functions,
converting recursive calls to loops, and
creating a huge \cc{main} function.
This, however, is
impractical for any reasonable-sized
real-world program
as analyzing a huge function
can be both time- and memory-intensive
in current compilers
while aggressive inlining 
results in large binaries.
Distinguishing heap allocations by thread
(i.e., thread-sensitivity) at compile-time
further adds complexity,
as it requires more extensive function cloning
to differentiate per-thread code statically.

Fortunately,
\sema can be deduced at runtime as well,
at the cost of code instrumentation
(and hence, overhead).
More specifically,
the dynamic deduction of \sema can be
facilitated with context-tracking logic
automatically and strategically
instrumented at compile-time.

\begin{figure}[t]
    \centering
    \small
    \hfill
    \hfill
    \begin{minipage}[c]{0.475\linewidth}
        \includegraphics[width=.9\linewidth]{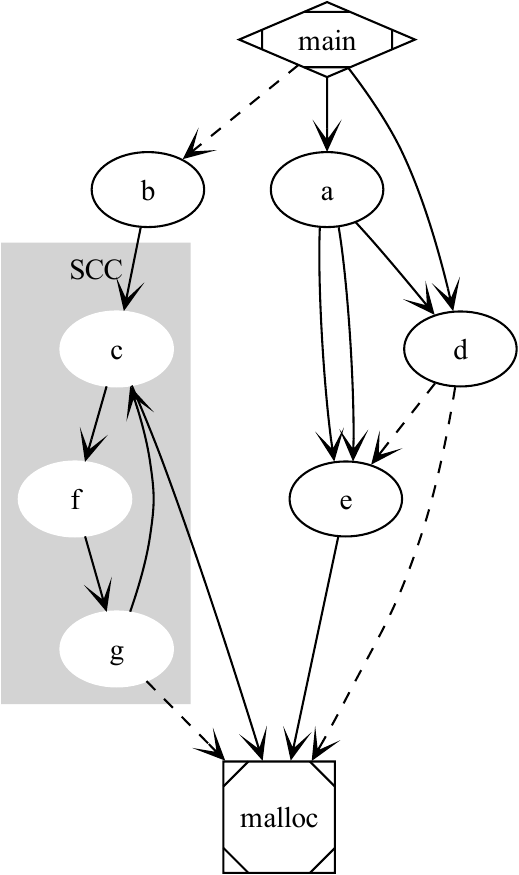}
    \end{minipage}
    \hfill
    \vline
    \hfill
    \begin{minipage}[c]{0.5\linewidth}
    \sema without recursive calls
    \begin{itemize}[noitemsep,topsep=0pt,leftmargin=15pt]
        \item[A] \cc{main}$\rightarrow$\cc{a}$\rightarrow$\cc{e}[l]$\rightarrow$\cc{malloc}
        \item[B] \cc{main}$\rightarrow$\cc{a}$\rightarrow$\cc{e}[r]$\rightarrow$\cc{malloc}
        \item[C] \cc{main}$\rightarrow$\cc{a}$\rightarrow$\cc{d}$\rightarrow$\cc{malloc}
        \item[D] \cc{main}$\rightarrow$\cc{a}$\rightarrow$\cc{d}$\rightarrow$\cc{e}$\rightarrow$\cc{malloc}
    \end{itemize}
    
    \vspace{-8pt}
    \hrulefill
    \vspace{-3pt}

    \sema with recursive calls
    \begin{itemize}[noitemsep,topsep=0pt,leftmargin=15pt]
    
        \item[E] \cc{main}$\rightarrow$\cc{b}$\rightarrow$\cc{c}$\rightarrow$\cc{malloc}
        
        \item[F] \cc{main}$\rightarrow$\cc{b}$\rightarrow$\cc{c}$\rightarrow$\cc{f}$\rightarrow$\cc{g}$\rightarrow$\cc{malloc}
        
        \item[G] \cc{main}$\rightarrow$\cc{b}$\rightarrow$\cc{c}$\rightarrow$\cc{f}$\rightarrow$\cc{g}$\rightarrow$
        \cc{c}$\rightarrow$\cc{malloc}
        
        \item[H] \cc{main}$\rightarrow$\cc{b}$\rightarrow$\cc{c}$\rightarrow$\cc{f}$\rightarrow$\cc{g}$\rightarrow$
        \cc{c}$\rightarrow$\cc{f}$\rightarrow$\cc{g}$\rightarrow$\cc{malloc}

        \item[I] \cc{main}$\rightarrow$\cc{b}$\rightarrow$\cc{c}$\rightarrow$\cc{f}$\rightarrow$\cc{g}$\rightarrow$
        \cc{c}$\rightarrow$\cc{f}$\rightarrow$\cc{g}$\rightarrow$\cc{c}$\rightarrow$\cc{malloc}

        \item[$\cdot$] $\cdots\cdots\cdots$
        \item[$\cdot$] $\cdots\cdots\cdots$
        
        \item[*] \cc{main}$\rightarrow$\cc{b}$\rightarrow$\cc{c}$\rightarrow$$\cdots$$\rightarrow$\cc{malloc}
    \end{itemize}
    
    \end{minipage}
    
    \caption{Call graph (left) of a crafted program~\autoref{app:code_example}
    illustrating how \sema (right) can be deduced.
    \textnormal{In this call graph, each node is a function and
    solid edges represent function calls
    not in a loop inside the corresponding function CFG while
    dashed edges represent function calls inside a loop.}}
    \label{fig:example}
\end{figure}

\PP{A concrete example}
To illustrate how \sema can be deduced,
we use the simple example in~\autoref{fig:example}.
The code snippet is shown in~\autoref{app:code_example}
and the figure is a conventional call graph of the program
enhanced with
(1) flow-sensitive edges
(e.g., two edges from \cc{a} to \cc{e}
marked as \cc{[l] and \cc[r]} respectively) and
(2) annotations on whether the call occurs inside a loop or not
(i.e., dashed vs solid edges).

\subsection{Cyclic Control-flow Structures}
\label{ss:design_cyclic}

Due to the existence of
a recursive call group (the SCC),
there are an unlimited number of call traces
that can reach \cc{malloc} from \cc{main}.
This is why
we cannot enumerate all call traces to
assign each call trace a \sema statically.
And yet,
even we can track the call context at runtime,
having an unlimited number of \sema{s} for this program
is not desirable either,
because such an approach is,
in the worst case,
the same as giving
each allocated heap object a different \sema.
As discussed in~\autoref{ss:bg-type},
this can be overly sensitive and
may cause a significant memory overhead
as in FFMalloc~\cite{ffmalloc}.

\emph{How can we fit unlimited number of call traces
into a fixed number of \sema{s}}?
We considered two simple solutions:

\begin{itemize}[noitemsep,topsep=0pt,leftmargin=*]
  \item \emph{Bounded unrolling}:
  unroll the SCC to a limited depth and
  treat each \cc{malloc} called from the unrolled
  iterations differently.
  Beyond the unrolled iterations,
  assign a single \sema to the \cc{malloc} called inside this SCC.

  \item \emph{Aggregation-based hitmap}:
  aggregate the call trace inside the SCC
  to a fixed number of bits;
  call traces with the same aggregated value
  are deemed to have
  the same \sema.
\end{itemize}

\noindent
\sys uses the aggregation-based hitmap solution
as it provides slightly better security
by distributing \sema more uniformly across
different rounds of recursion.

However,
\cc{malloc} calls
occurring in different loop iterations
are not differentiated by \sema.
Differentiating loop iterations
will require path-sensitive instrumentation,
i.e., instrumentations (and hence overhead)
linear to the number of basic blocks;
while differentiating iterations
in recursive calls only requires
instrumentations linear to the number of functions,
which is arguably significantly smaller 
in most real-world programs.
This helps to reduce the performance impact
caused by instrumentation.

\ignore{
In this example,
five \sema are involved:
\begin{enumerate}
    \item \cc{main} $\rightarrow$ \textbf{D} $\rightarrow$ \cc{malloc(sizeof(int))};
    \item \cc{main} $\rightarrow$ \textbf{B} $\rightarrow$ \textbf{D} $\rightarrow$ \cc{malloc(16)};
    \item \cc{main} $\rightarrow$ \textbf{B} $\rightarrow$ \cc{malloc(16)}
    \item \cc{main} $\rightarrow$ \textbf{C} $\rightarrow$ \textbf{D}$\rightarrow$ \cc{malloc(16)}
    \item \cc{main} $\rightarrow$ \textbf{C} $\rightarrow$ \textbf{D}$\rightarrow$ \cc{malloc(sizeof(int))}
\end{enumerate}

However, among these five \sema,
only the first three \sema need to be tracked,
and the last two \sema shall only be associated with at most one block allocation.
For performance consideration,
we only assign weights for edges that
call or eventually call a heap memory allocation function.
In this example,
the first three \sema are assigned with 
the \nid: 0, 1, and 2 respectively.

Reflecting this example,
TypeAfterType takes the first and the last 
\sema as the same ``type'' as they allocate
blocks with the size traced back to the same 
object (\cc{int}).
It also takes the second and the forth \sema
as the same,
as no \cc{sizeof} tracing information is available,
while allocated on the same final call site.
Cling also takes them as the same,
as they have the same call depth and
the final \cc{malloc} are all 
called on the same site.
}

\subsection{\sema Representation}
\label{ss:overview-repr}

\sema can be represented as
a composition of two values:
\begin{itemize}[noitemsep,topsep=0pt,leftmargin=*]
\item \nid: a non-recurrence identifier
    representing top-level call traces
    in the \emph{directed acyclic} call graph,
    which is built by
    abstracting each SCC in the call graph into a node;

\item \rid: a recurrence identifier
for call traces within an SCC.
\end{itemize}
\vspace{2pt}

\noindent
The \nid and \rid for the current execution context
are both tracked at runtime through global variables.
Their values are merged together to form a \sema
when the execution is about to invoke
a memory allocation function (e.g., \cc{malloc}).

We assign each call site outside SCCs with a \emph{weight} (\autoref{ss:design_weight}).
Before making a call,
\nid is incremented by the weight of the call site and 
decremented by the same weight upon return.
This rule for \nid generalizes
to a stack of calls as well.
Operationally,
\nid is the cumulative weight of all call sites
in the call stack when a heap allocation happens.
Our weight assignment algorithm (\autoref{ss:design_weight})
ensures that two \sema instances
have the same \nid if and only if their external SCC traces are identical
(formally proved in~\autoref{app:sec_formal}).

\rid is for intra-SCC call stack tracking.
Unlike \nid,
\rid is an aggregated value of what happened inside an SCC.
\rid is tracked with
two global variables $s$ and $h$, where

\begin{itemize}[noitemsep,topsep=0pt,leftmargin=*]
\item $s$ is a stack that hosts the
stack pointers before a function within an SCC is called
(a.k.a., a call stack), and

\item $h$ is the aggregation of stack $s$,
representing the \rid~(\autoref{ss:design_hash}).
$h$ is computed and stored before an SCC function 
calls a function not in the current SCC
(an outbound call),
and is cleared after the call to this SCC
(an inbound call) returns.
\end{itemize}

\PP{Repetitive allocation}
A \sema only needs to be tracked if
heap objects of this \sema can be allocated repetitively.
For one-time allocations,
i.e., a \sema that can only be
reached in one call stack where
none of the call site is in a loop
(see~\autoref{ss:eval-other-stats}
for evidence that this is rare),
once an one-time object is freed,
its space is never reused.
Therefore,
we optimize \sema tracking
only to those instances where re-allocation is possible,
identified by the presence of at least
one recursive call site in their allocation traces.

We keep track of the recursive depth $l$
by incrementing it before 
executing an iterative function call
and decrementing after it.
Upon memory allocation, if $l$ is not zero,
we can conclude that this \sema object may be  
recurrently allocated.
$l$ is also increased before a non-SCC
function calls a SCC function (inbound call)
and decreased after it.

\PP{Illustration}
Revisiting the example in~\autoref{fig:example},
 only \textbf{A} and \textbf{B}
are non-repetitive; 
all other \sema{s} need to be tracked:
types \textbf{C} and \textbf{D} are repetitive
because of looping while
other types are repetitive due to involvement in
recursive calls.

We take \textbf{F} as a case study for variable management.
The \nid
is increased before calling \cc{c} and \cc{malloc}.
The call stack $s$ holds three stack pointers,
pushed into it before each 
 SCC function (\cc{c}, \cc{f}, and \cc{g}) is called.
Upon calling \cc{malloc}, the \rid
(i.e., $h$) is computed.
The \cc{b$\rightarrow$c} function call enters a SCC, causing
 $l$ to be incremented.
Upon calling \cc{malloc},
$l$ is non-zero 
indicating a recurrent allocation.

\PP{Thread sensitivity}
Note that thread identifiers are not
discussed here in the representation of \sema
despite the fact that \sema is thread-sensitive.
This is because
the backend heap allocator does not need
this information to be deduced through
compiler-instrumented code at runtime.
Instead,
it can be queried directly by the
backend allocator via
a system call (e.g., \cc{syscall(\_\_NR\_gettid)})
or even one assembly instruction
if the platform supports.
As a result,
we do not specifically encode a thread ID
in the \cc{malloc} argument passed to the
backend allocator (see~\autoref{ss:design_encode}).

\subsection{Alternative: Path-sensitivity}
\label{ss:overview-path}

\sema is not path-sensitive.
Although a
\emph{thread-, context- and path-sensitive
type qualifier} is intriguing,
we have to weaken path- to flow-sensitivity
for practical reasons:
\begin{itemize}[noitemsep,topsep=0pt,leftmargin=*]
    \item Within a function control-flow graph (CFG),
    paths exponentially outnumber CFG nodes
    (the latter is captured by flow-sensitivity),
    hence adopting a path-based \sema will bloat
    the number of \sema{s} and allocation pools.
    
    \item Deducing execution paths requires either
    dynamic CFG branch tracking (non-trivial run-time overhead) or
    static function splitting,
    e.g., assign different \sema{s} to the same \cc{malloc}
    based on whether function arguments satisfies predicate $X$,
    except that devising $X$ is undecidable.

    \item Empirical evaluation (\autoref{s:sec}) shows that
    \sema in its current form
    is sufficient to defend against known exploits.
\end{itemize}

\section{\NoCaseChange{\sema}-based Heap Allocation}
\label{s:design}

In this section,
we describe the design and implementation details of \sys---a
\sema-based heap allcator for mitigating UAF vulnerabilities.
We first introduce our threat model and explain
how \sys realizes dynamic \sema deduction and 
allocates memory accordingly.

\PP{Threat model}
We assume that
(a) the underlying operating system kernel and hardware are trusted,
(b) the targeted program is uncompromised at startup, and
(c) the attacker can
obtain and analyze the source code and
the compiled binaries of both
the targeted program and \sys.
Exploiting implementation bugs
in \sys 
or utilizing side-channel information
(e.g., cache and power usage)
is out of scope.

\subsection{Overview}

\sys consists of an LLVM transformation pass
and a heap allocator backend.
The LLVM pass analyzes the 
intermediate representation (IR),
inserts instructions to create
and instrument the tracking variables,
and encodes the allocation parameters 
with \sema-tracking information.
The LLVM pass is built on top of
MLTA~\cite{mlta}
(for comprehensive and robust call graph construction)
and CXXGraph~\cite{cxxgraph}
(for graph algorithms).
The pass instruments
\sema tracking and encoding in the program which
eventually passes \sema
through common heap allocation APIs (see~\autoref{app:support_api} for a list).
The heap allocator backend 
takes the encoded information
for segregated memory allocation.

\begin{figure}[t]
    \centering
    \includegraphics[width=.40\textwidth]{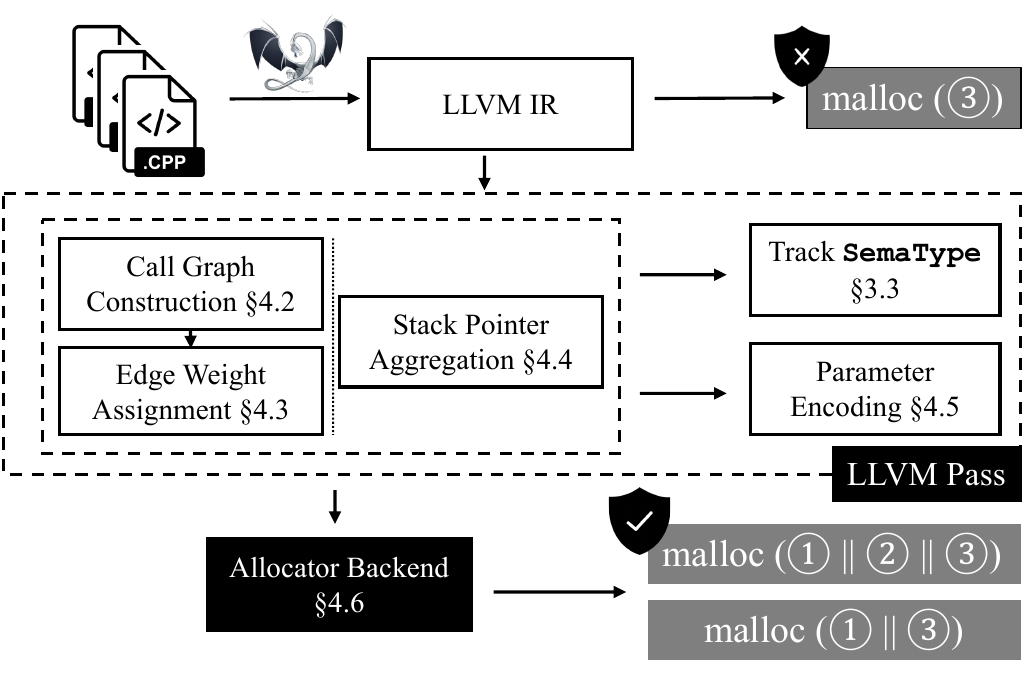}
    \caption{Design overview of \sys (\protect\WC{1}: flags, \protect\WC{2}: \sema, \protect\WC{3}: allocation size). The size is the parameter without \sys, while \sys encodes the trace information into the parameter after applying the pass.}
    \label{fig:flow}
\end{figure}

\autoref{fig:flow} gives a comprehensive overview of \sys.
In the transformation pass, 
\sys first constructs a call graph that 
only contains
functions (nodes) and 
call sites (edges)
relevant to \sema that need to be tracked (see~\autoref{ss:design_identify}),
and assigns weights on all edges and nodes in it
for \nid computation
(see~\autoref{ss:design_weight}).
In the call graph,
an SCC is treated as a function node,
and call traces within it are not considered by \nid.
Instead, intra-SCC calls are tracked in \rid 
by obtaining and aggregating the stack pointers
with an aggregation algorithm (see~\autoref{ss:design_hash}).
They are encoded into the \cc{size}
parameter of an allocation request (see~\autoref{ss:design_encode}).

We explain how the allocator backend
enforces allocation segregation using \sema in~\autoref{ss:design_enforce}.
For simplicity,
we use \cc{malloc} 
to represent all functions that may
request heap memory \emph{directly} from the backend.
We refer readers to~\autoref{app:insertion}
for a complete discussion about how the IR is 
transformed after applying the pass and 
how instructions are inserted.

\subsection{Call Graph Construction}
\label{ss:design_identify}

We start by building a call graph
for the program to be hardened by \sys.
While call graph is a foundational concept
with mature support in modern compilers,
the call graph in \sys is slightly more complicated
in two aspects:

\textbf{1) Flow-sensitive edges.}
If function \cc{e} is called in two places
by function \cc{a},
there will be two edges from \cc{a} to \cc{e}
in the call graph,
as shown in the example in~\autoref{fig:example}.

\textbf{2) Indirect calls.}
\sys takes special care for indirect calls
whose call targets cannot be resolved at compile-time
and hence do not show up in a conventional call graph.
To handle indirect calls,
\sys first identifies all callee candidates
via MLTA~\cite{mlta},
i.e., by matching the function type hierarchically. 
Subsequently,
for each callee candidate identified,
\sys adds an edge in the call graph
and treat different callee candidates
as if they are called in different places
in the calling function.
This is a conservative treatment for indirect calls and
can lead to more \sema{s} being derived than necessary
which can result in
a better security but
a larger memory overhead.

\PP{Additional trimming and marking}
With a baseline call graph,
the next step is to
remove nodes and edges that are irrelevant to \sema,
i.e., paths that do not eventually lead to a
\cc{malloc}.
We also mark call sites that occur
in a loop in the caller function
(e.g., dashed edges in~\autoref{fig:example})
in order to distinguish recurrent allocations
vs one-time allocations (see~\autoref{ss:design_cyclic}).
This marked call graph enables \sys to
optimize instrumentation to
recurrent \cc{malloc}s only.
We remove all nodes or edges that are not (eventually) called by or (eventually) call any recurrent edge.
This call graph contains only nodes and edges that 
eventually call \cc{malloc} while each edge leads to 
at least one recurrent \sema object.

Finally, we use the
Kosaraju-Sharir algorithm~\cite{kosaraju} to 
identify SCCs and
create a new call graph with each SCC being
abstracted as a single node.
In this way, the new call graph is essentially
a directed acyclic graph (DAG)
while recursions (i.e., intra-SCC paths)
are handled using \rid (see~\autoref{ss:design_hash}).

\subsection{Edge Weight Assignment}
\label{ss:design_weight}

Recall from~\autoref{ss:overview-repr}
that \nid,
which is part of the representation for \sema,
serves to distinguish different call stacks
that end up with \cc{malloc}
modulo recursions in SCCs.
As \nid is calculated as the sum of weights
per each edge in the path,
these weights need to be assigned
strategically to ensure that
different paths yields different \nid values.

To assign weights,
we run a topological sort on the DAG for
a deterministic ordering of functions
and then go through each function to assign weights
according to~\autoref{alg:weight_assign}.

\begin{algorithm}
\caption{Edge weight assignment.}
\label{alg:weight_assign}
nodes $\leftarrow$ topological\_sort(DAG) \\
\For{\textnormal{\textbf{each} n $\in$ nodes}}{
    w $\leftarrow$ 0 \\
    \For{\textnormal{\textbf{each} e $\in$ n.outgoing\_edges}}{
        e.weight $\leftarrow$ w \\
        w $\leftarrow$ w + max(1, e.dst.weight) \\
    }
    n.weight $\leftarrow$ w
}
\end{algorithm}

We maintain two weights while going through each function:
the function weight and the call-site weight.
The function weight describes
how many different \sema{s} exist 
if taking this function as the program entry point.
The call-site weight is the sum of the weights
of all functions called before it within the function.
It describes how many different \sema{s}
all previous call sites of the current function lead to.
More specifically, 
it is an offset that guarantees that all \sema{s} allocated
through the current call site have their \nid larger than 
all previous \sema \nid{s} to avoid collision.
For example, in a function, the path weight 
of the first call site is zero,
and the weight of the next call site 
is the weight of the first callee function
(note the minimum weight is one and
\cc{e.dst} is a node whose
weight has been computed in a previous round,
line 4--7).
As long as the offset is computed correctly,
no collision will happen.

After processing all call sites, 
we assign the weight of the current 
function as the sum of the weights
of all its callees (line 8).
Using the topological order,
we guarantee that all callee 
weights are computed before they are needed;
we set the weight of \cc{malloc} 
to zero as it does not call any function.

Note that weight assignment (\autoref{alg:weight_assign})
ensures a one-to-one mapping between
a \nid and
an \textbf{end-to-end path} that reaches the \cc{malloc} in the call DAG
(see~\autoref{app:sec_formal} for a proof).
It is worth-noting, however, that
tracking the path at runtime directly is possible
but would incur a slightly higher overhead than
tracking the \nid,
which only involves two arithmetic operations
per each call site.

\PP{Optimization}
To further minimize the instrumentation needed,
a node can be removed from the call graph 
if it has only one incoming edge,
i.e., the function \cc{f} (represented by this node)
is only called in one place.
Essentially,
removing the node has the same effect as
inlining \cc{f} into its caller
(without actually transforming the code).
In this situation,
the call site that invokes \cc{f} does not need to be
instrumented for \nid-related logic.
And this optimization repeats until
we cannot find such \cc{f} in the call graph.

\subsection{SCC Stack Pointers Aggregation}
\label{ss:design_hash}

We
use an aggregation approach to track the execution
path within SCCs.
Before calling each function within the SCCs,
we obtain and push the stack pointer into the stack $s$,
which is the aggregation input to compute \rid.

\begin{algorithm}
\caption{Aggregation of stack pointers.}\label{alg:hash_stack}
h $\leftarrow$ 0\\
\For{\textnormal{\textbf{each} p $\in$ s}}{
    h $\leftarrow$ h $\ll$ 2 ; \,\,
    p $\leftarrow$ (p $\gg$ 6) \cc{\&} \cc{0x3} ; \,\,
    h $\leftarrow$ h + p
}
h $\leftarrow$ h \cc{\&} \cc{0x3FFFF}
\end{algorithm}

\rid is computed using~\autoref{alg:hash_stack}.
Initially, we set it to be zero (line 1).
We then go through each stack pointer 
by adding 7th and 8th least significant bits (LSBs)
of each input to it and
shift it left by two bits (line 3-5).
We specifically take these two LSBs
as stack pointers are 
8-byte aligned in the x86 clang environment~\cite{llvm-align},
and we select those bits that are not identical in 
different call frames.
Finally,
we only keep the least fourteen bits of the aggregated value,
which represents the most recent seven functions called within SCCs. 

We note that as a stack pointer is dependent on the call depth
and all calls that are not returned,
this algorithm accounts for the entire call trace 
without losing function calls older than the most recent seven.
The current parameter maximizes tracking depth with 
a minimal 2-bit entropy to differentiate call frames.
If recursive calls are shallow or many functions are involved in an SCC, 
it will be preferable to track fewer layers
with more bits taken per pointer.

\subsection{Parameter Encoding}
\label{ss:design_encode}

The heap allocator backend
requires two pieces of information as input:
allocation size (as required by all memory allocators)
and \sema (unique information in \sys),
and allocates heap objects based on them.
While standard memory allocation APIs
already accept the allocation size as a parameter,
we need to find a way to pass \sema to the backend.
And \sys, conceptually, has two options:

\begin{itemize}[noitemsep,topsep=0pt,leftmargin=*]
  \item Changing the \cc{malloc} signature:
  This would involve
  adding a new parameter to the existing interface and hence,
  introducing a new function signature like
  \cc{malloc(size\_t size, void *semantics)}.
  
  \item Repurposing the \cc{size\_t} parameter type:
  This implicitly change the type of the \cc{size} parameter
  with \sema encoded alongside the existing size. 
\end{itemize}

\begin{figure}
    \centering
    \includegraphics[width=.47\textwidth]{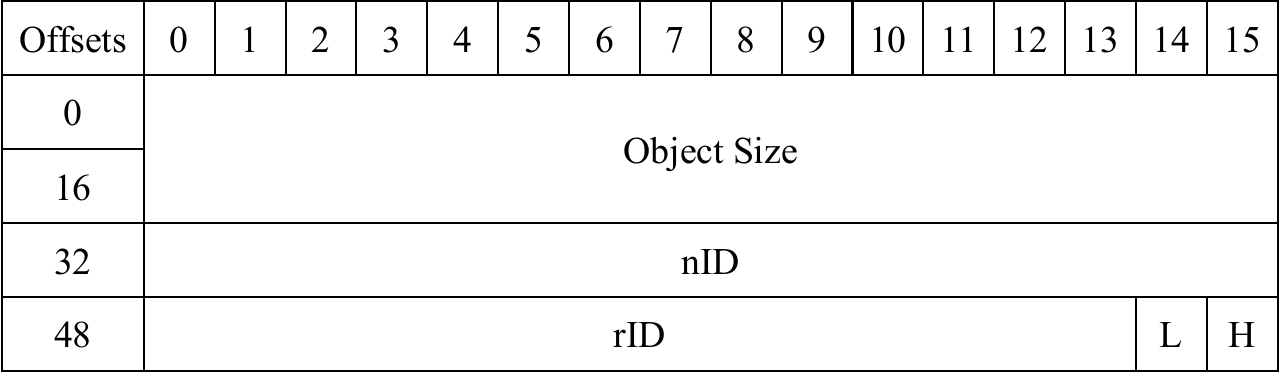}
    \caption{Parameter encoding rule for regular objects(L: loop identifier; H: huge block identifier).}
    \label{fig:encoding}
\end{figure}

\noindent
In \sys, we take the second approach 
for compatibility with the existing allocation interface,
and encode \sema within the 
\cc{malloc} \cc{size} parameter using the format
shown in~\autoref{fig:encoding} for blocks smaller than 4GB.

We set the loop bit (L) if the number of loop 
layers ($l$) is not zero to notify the backend that it might
reuse the memory freed by another object.
We store the \nid and \rid accordingly
as the \sema,
and we use the remaining 32 bits to store the size of the allocated object.

For larger blocks, 
we set the huge-block bit (H) and use all the 
remaining 63 bits to store the block size.
These blocks are allocated via a system call,
and launching UAF attacks on them is not 
trivial (see~\autoref{app:backend} for details).

This design is compatible with 
legacy code or external libraries that
are not transformed by our pass with 
function allocation call size up to 4GB.
However, memory allocated this way 
does not have the loop identifier set,
and is not going to be released
unless the block is big enough that allocated
from the OS directly (see~\autoref{app:backend} for details).
This indeed is not common as shown in~\autoref{tab:allocations},
where most tests have more than 99\% allocations identified
with recurrent \sema.

\PP{Tunable parameters}
While Loop (L) and huge (H) indicators are both 1-bit and are not tunable,
other parameters can be tuned to fit specific program or security requirements.

\begin{itemize} [noitemsep, topsep=0pt,leftmargin=12pt]
    \item \emph{\cc{size}} (default 32-bit):
    A smaller \cc{size} leaves more room for \nid and \rid,
    but may impair functionality:
    if legacy code or
    external libraries that are not transformed by \sys
    incur larger allocations,
    the overflowed bits will be taken as \sema,
    causing the allocated block to be smaller than required.
    The 32-bit default is an empirical number
    based on programs we evaluated.

    \item \emph{\nid} (default 16-bit):
    Ideally, the size of \nid should be the smallest number that satisfies
    $2^{sizeof(nID)}\ge\sum P_i$ for a target program,
    where $P_i$ represents the number of traces
    to reach an allocation site $i$
    in the flow-sensitive call graph after DAG-reduction
    (\autoref{ss:design_identify}).
    We take the 16-bit default to support all tested programs and benchmarks (see~\autoref{s:eval}).

    \item \emph{\rid} (default 14-bit):
    Given a 64-bit pointer,
    the size of \rid is passively decided by \cc{size} and \nid. 
    However, the calculation of \rid is tunable, 
    as discussed in~\autoref{ss:design_hash}.
\end{itemize}

\subsection{Heap Allocator Backend}
\label{ss:design_enforce}

The backend heap allocator for \sys is packaged
as a library that can either be
preloaded at loading time or
statically linked
to replace the default allocator.
The backend extracts and decodes the \sema
packed in the \cc{size} parameter and
enforce \sema-based allocation by
allocating objects of different \sema{s} from
segregated pools.

More specifically,
\sys backend adopts BIBOP~\cite{bibop} for
block allocation inside each \sema pool.
BIBOP allocates blocks of the same size class together 
using one or more continuous memory pages, and 
preemptively allocate sub-pools for each size class.
A block is not going to be further split or coalesced.
\sys is built upon this design.
For each thread (hence thread-sensitivity~\autoref{ss:bg-type})
it allocates
a global BIBOP pool for one-time \sema{s} and
individual pools for different \sema{s}
upon seeing a recurrent request for the same \sema.
\sys uses power of two size classes,
for example, all blocks
with the same \sema and
of size 65 to 128 bytes will be allocated to the same pool.

Operationally,
upon receiving a heap allocaiton request,
the backend first
checks the huge bit and, if applicable,
allocates huge blocks using
the \cc{mmap}~\cite{mmapzero} system call.
For regular blocks, 
if the loop bit is not set,
\sys will allocate it using the global pool,
and it will never be released even after it is freed.
If the loop bit is set, 
\sys allocates it using the global pool
if the \sema is seen for the first time and
otherwise create an individual pool dedicated
for all following allocations
with this \sema
A freed block in the individual pool
can be reused by later allocations with the 
same \sema.
We refer readers to~\autoref{app:backend}
for additional implementation details
of this runtime backend.
\section{Security Analysis}
\label{s:sec}

We provide both qualitative analysis and empirical evidence
on the effectiveness of \sys in mitigating UAF exploits.

\subsection{Qualitative Analysis}
\label{ss:sec_formal}

The key reason why type-based allocators
cannot deliver perfect UAF mitigation is
UAF within the same type.
More specifically, to \sys,
this means UAF within memory objects
marked with the same \sema---and
this is not only possible but also common
due to recurrent allocations,
i.e., \cc{malloc} inside a cyclic control-flow structures
such as loops or recursive calls
(\autoref{ss:design_cyclic}).
On the other hand,
memory reuse is crucial in reducing memory footprint.
An allocator that places each object into a new pool and
never reclaim memory is immune to UAF at the cost of
a high memory waste.
Therefore, intuitively,
the more recurrent allocations a program have,
the less effective  \sys is in mitigating UAF exploits,
but the greater the memory saved by \sys,
compared to allocators that never free memory.

In this section,
we sketch a qualitative explanation on
how loops and recursive calls 
affect the security of \sys.

\PP{Setup}
Assume a program has $N$ allocation sites:
\begin{itemize}[noitemsep,topsep=0pt,leftmargin=*]
    \item each allocation site $i \in 1..N$ can be reached via
    $P_i$ traces in the flow-sensitive call graph
    after CFG-reduction (\autoref{ss:design_identify});

    \item each trace $T_{i,j}$ (where $j \in 1..P_i$)
    contains $R_{i,j}$ nodes
    that are reduced from call graph SCCs,
    i.e., recursive calls (\autoref{ss:design_hash});

    \item $k$ out of $N$ sites are in a loop
    w.r.t a function-level CFG.
\end{itemize}

\vspace{2pt}
\noindent
\sys assigns one \nid to each trace $T_{i,j}$ 
and up to $2^{\text{\#bits}(rID)}$ \rid{s} per trace. 
Thus,
this program will have:
\begin{itemize}[noitemsep,topsep=0pt,leftmargin=*]
\item a minimum of $\sum P_i$ \sema{s},
the minimum occurs when the program
does not contain any recursive calls, or

\item a maximum of $2^{\text{\#bits}(rID)} \times \sum P_i$ \sema{s},
the maximum occurs when
all traces $T_{i,j}$ have recursive calls.
\end{itemize}

\PP{UAF-protection in different scenarios}
We discuss how UAF protection in \sys
can be weakened by recurrent allocations
with reference to
complete UAF mitigation:

\begin{itemize} [noitemsep, topsep=0pt,leftmargin=12pt]
    \item \emph{No recurrent allocation}
    ($k=0$ and all $R_{i,j}=0$):
        \sys provides perfect security against UAF,
        a similar level of protection as
        complete UAF-mitigating allocators
        since no memory reuse exists.
        \sys provides strictly more protection
        than existing type-based allocators:
        in the worst case,
        all $\sum P_i$ \sema{s} can the same C/C++ type
        which will be allocated from the same pool.

    \item \emph{With \cc{malloc} in loops}
    ($k\ne0$ and all $R_{i,j}=0$):
        \sys provides weaker security than
        complete UAF-mitigating allocators
        as UAF is possible within the same \sema
        in one of the $k$ loop allocations.
        Higher $k$ means weaker UAF protection, 
        but a smaller memory footprint. 
        Regardless of $k$,
        \sys provides strictly more protection
        than existing type-based allocators,
        as shown in~\autoref{ss:bg-type}.

    \item \emph{With \cc{malloc}
    in one group of recursive calls only}
    ($k=0$ and all $R_{i,j}=0$ except $R_{a,b}=1$):
        \sys provides weaker security than
        complete UAF-mitigating allocators
        as UAF is possible among the same \sema
        in the recursive call group. 
        Every \sema in trace $T_{a,b}$ shares the same \nid, 
        and there is only limited entropy for \rid 
        but potentially unlimited call traces
        in the recursive call group.
        Having more SCCs in call graphs means weaker security
        but more memory-saving. 
        Regardless of the number of SCCs, 
        \sys still provides strictly more protection
        than existing type-based allocators,
        as shown in~\autoref{ss:bg-type}.
        
    \item \emph{With \cc{malloc}
    in both loops and recursive calls}
    ($k\ne0$ and some or all $R_{i,j}\ne0$):
    Security degradation comes from
    all sources of recurrent allocations
    (discussed above)
    as there are now more chances for
    two objects to be marked as the same \sema.
    However,
    memory savings are also brought in
    due to exactly the same reasons.
\end{itemize}

\PP{Effectiveness evaluation}
To show how \sema diversifies heap allocation,
we compare the number of different allocation sites using 
\sema{s} and pure object types in the 
last two columns of~\autoref{tab:allocations}
based on programs in
the PARSEC3~\cite{parsec} and
SPEC 2017~\cite{spec} benchmarks.
In programs that have complicated program contexts
(e.g., 600 and 602),
\sema diversifies the allocations by more than 250x than 
the native allocation sites.
Other tested programs that have the same native-typed objects 
allocated from different traces are also diversified accordingly.

\subsection{Empirical Check on Real-world Exploits}
\label{ss:sec_cves}

We evaluate the effectiveness of
\sys in stopping UAF exploits 
by running it with 15 real-world UAF vulnerabilities.
We compare the protection results
with two type-based allocators,
Cling~\cite{cling} and
TypeAfterType~\cite{typeaftertype},
while other allocators
used in performance evaluation (\autoref{s:eval})
either have theoretically complete UAF-mitigation~\cite{dangzero, markus, minesweep, ffmalloc}
or requires case-by-case manual annotation to work
(e.g., PUMM~\cite{pumm}).

Tested vulnerabilities are summarized 
in~\autoref{tab:sec_cves}.
They are selected from three sources:
representative CVEs from
DangZero~\cite{dangzero},
TypeAfterType~\cite{typeaftertype},
uafBench~\cite{uafbench},
and further
enriched with additional vulnerabilities selected 
by us to cover the exploitation types discussed
in~\autoref{ss:bg-exploit}.
We present two representative examples here
and two more in~\autoref{app:additional_cve}.

While \sys successfully thwarts all exploits,
TypeAfterType provides no defense against four exploits
and only partial protections for most attacks,
as the attacker can still launch attacks successfully but
cannot create powerful attack primitives.
Additionally,
we checked all exploits since 2019 in
exploitDB~\cite{exploitdb},
and we are not aware of any exploitation against \sys---confirming
that \sys can help confine UAF exploitability in practice.

\begin{table}[t]
    \centering
    \small
    \begin{tabular}{lcccc}
    \toprule
        \textbf{Vulnerability} & 
        \textbf{Exp.} (\autoref{ss:bg-exploit}) & \textbf{\cite{typeaftertype}} & 
        \textbf{\cite{cling}$^\dagger$} &
        \textbf{\sys} \\ \midrule
        CVE-2015-6831 & B & \Circle & \Circle & \CIRCLE \\
        CVE-2015-6835  & C & \Circle & \Circle & \CIRCLE \\
        Python-24613& C & \CIRCLE & \CIRCLE & \CIRCLE \\
        mRuby-4001 & D & \LEFTcircle & \LEFTcircle & \CIRCLE \\
        yasm-91 & D/E & \LEFTcircle & \Circle & \CIRCLE \\
        CVE-2018-11496 & D/E & \Circle & \Circle & \CIRCLE \\
        CVE-2018-20623 & C & \CIRCLE & \LEFTcircle & \CIRCLE \\
        yasm-issue-91 & C & \LEFTcircle & \LEFTcircle & \CIRCLE \\
        mjs-issue-78 & B & \LEFTcircle & \Circle & \CIRCLE \\
        mjs-issue-73 & B & \LEFTcircle & \Circle & \CIRCLE \\
        CVE-2017-10686 & D/E & \LEFTcircle & \Circle & \CIRCLE \\
        CVE-2016-3189 & D & \Circle & \Circle & \CIRCLE \\ 
        CVE-2009-0749 & D/E & \CIRCLE & \CIRCLE & \CIRCLE \\
        CVE-2011-0065 & B & \CIRCLE & \LEFTcircle & \CIRCLE  \\
        CVE-2012-0469 & B & \CIRCLE & \LEFTcircle & \CIRCLE  \\
        \bottomrule
    \end{tabular}
    \caption{
    \sys is effective in thwarting (\CIRCLE) exploitation of all
    real-world UAF vulnerabilities evaluated
    while TypeAfterType~\cite{typeaftertype} and Cling~\cite{cling}
    provide no protection (\Circle) or
    partial protection (\LEFTcircle) to most vulnerabilities.
    $^\dagger$: Cling is not open-sourced and
    is only analyzed conceptually.}
    \label{tab:sec_cves}
\end{table}

\noindent
\textbf{Case study: mjs-issue-78}~\cite{uafbench}.
This vulnerability is in \cc{mjs},
a restricted JavaScript engine,
and can be triggered when
\cc{mjs} parses a crafted \cc{JSON} string as shown in \href{https://github.com/cesanta/mjs/blob/238dc31c6eb386bd91f3a3f1491fc46b650783b1/mjs/tests/unit_test.c#L2838-L2842}{the test case}.

While parsing,
both the raw \cc{JSON} string and intermediate outputs
are stored in one buffer:
field \href{https://github.com/cesanta/mjs/blob/9eae0e6e8fbfa25b71ea2446d9ee667c5c4271fe/mjs.c#L3126}{\cc{owned\_strings}}
within type (\href{https://github.com/cesanta/mjs/blob/9eae0e6e8fbfa25b71ea2446d9ee667c5c4271fe/mjs.c#L2098-L2102}{\cc{struct mjs}}),
a context manager for an \cc{mjs} engine.
As the parser keeps appending parsed elements to the buffer
(more precisely, to \cc{mjs->owned\_string->buf})
during
\href{https://github.com/cesanta/mjs/blob/9eae0e6e8fbfa25b71ea2446d9ee667c5c4271fe/mjs.c#L13656}{\cc{mjs\_mk\_string}},
the buffer might potentially be reallocated
via \href{https://github.com/cesanta/mjs/blob/9eae0e6e8fbfa25b71ea2446d9ee667c5c4271fe/mjs.c#L13706}{mbuf\_resize},
causing other pointers
that also refer to the same buffer to be dangling
(e.g., \href{https://github.com/cesanta/mjs/blob/9eae0e6e8fbfa25b71ea2446d9ee667c5c4271fe/mjs.c#L6430}{\cc{frozen->cur}}).
To summarize,
the dangling pointer in this UAF vulnerability
is allocated in the call trace of
\cc{mjs\_json\_parse} $\rightarrow$
\cc{json\_walk} $\rightarrow$ \cc{...}  $\rightarrow$
\cc{frozen\_cb} $\rightarrow$
\cc{mjs\_mk\_string} $\rightarrow$
\cc{mbuf\_resize} $\rightarrow$
\cc{realloc}.

Assuming that the memory chunk freed by \cc{mbuf\_resize}
is later reallocated to a buffer in which
the attacker can put arbitrary data,
then the attacker-controlled object
can be accessed by a dangling pointer
(e.g., the \cc{frozen->cur}
through one of 
``\href{https://github.com/cesanta/mjs/blob/9eae0e6e8fbfa25b71ea2446d9ee667c5c4271fe/mjs.c#L5966}{\cc{cur(f)}}'').
This UAF-read
might lead to compromised execution states,
making it a type-B exploit.

From an attackers' perspective,
to exploit this UAF vulnerability,
the crux is to gain control of an object that
may be allocated to the memory chunk
freed by \cc{mbuf\_resize}.
This can be done in at least two ways
based on our findings:

\emph{Exploit 1}:
Run an \cc{mjs} engine in another thread
and have the other \cc{mjs} engine parse
an attacker-supplied \cc{JSON} string.
In this way,
the attacker-controlled buffer is allocated using
exactly the same call trace as the dangling pointer.
Therefore,
only \sys can defend against this exploit
because \sema is thread-sensitive
while flow- and context-sensitivity is not enough.

Recall that
Cling defines the ``type'' of an allocated object
based on the two innermost return addresses on the call
stack when \cc{realloc} is invoked.
This definition cannot distinguish objects allocated
using exactly the same call stack on different threads.
The lack of thread-sensitivity is also the reason
why TypeAfterType cannot defend against this exploit,
as both the freed object
(which inadvertently creates dangling pointers) and
attacker-controlled object
are classified as the same ``type'',
hence allowing UAF among them.

\emph{Exploit 2}:
An attacker may exploit another
call trace
\cc{mjs\_mkstr} $\rightarrow$
\cc{mjs\_mk\_string} $\rightarrow$
\cc{mbuf\_resize} $\rightarrow$
\cc{realloc}
to obtain a controllable buffer
potentially in the same thread where
\cc{mjs\_json\_parse} is invoked
(e.g., by placing a \cc{mkstr(..)} JavaScript call
after the \cc{JSON} string).
In this way,
the attacker-controlled buffer is allocated using
a different call trace as the dangling pointer.
\sys mitigates this exploit by assigning different
\sema{s} to the dangling pointer and attacker-controlled
buffer,
eliminating the possibility of UAF among them.

Cling takes \cc{mbuf\_resize} as an allocation wrapper
and treats all objects allocated through
\cc{mjs\_mk\_string} to have the same type.
This allows UAF between the dangling pointer
and attacker-controlled buffer despite that
they are originated from different roots.
TypeAfterType, on the other hand,
further takes \cc{mjs\_mk\_string}
as a \cc{malloc} wrapper
as it still passes a variable length to \cc{mbuf\_resize}.
This enables TypeAfterTypet to
differentiate objects allocated through
\cc{frozen\_cb} and \cc{mjs\_mkstr}.
Hence, can mitigate this exploit.

\noindent
\textbf{Case study: CVE-2015-6835}~\cite{CVE-2015-6835}.
This vulnerability is in the \cc{PS\_SERIALIZER\_DECODE\_FUNC}
function,
which restore a PHP session from a serialized string.
During this process,
\cc{php\_var\_unserialize} returns a \cc{zval} pointer,
which is stored in a hashtable.
However, the same pointer might be freed later and
this causes the stored copy to be dangling.
Through this dangling pointer,
an attacker might corrupt any \cc{zval} object
that may be reallocated to the freed slot.

\cc{zval} is a reference-counting wrapper of
almost all other objects in the PHP engine.
Therefore,
the attacker can corrupt any \cc{zval} object that may be reallocated to this free slot and
its value can be leaked through the dangling pointer.
In the PoC exploit,
the attacker simply uses the PHP \cc{echo(..)}
function to dump a newly allocated
\cc{zval} through the dangling pointer,
i.e., a type-C exploit.

In this PoC exploit, 
both the dangling pointer and
victom objects are allocated through
a common call trace:
\cc{php\_var\_unserialize} $\rightarrow$ \cc{emalloc} $\rightarrow$ \cc{malloc}.
This is critical to understand
why both Cling and TypeAfterType
fail to provide protection.
For Cling,
this \cc{malloc} wrapper chain implies that
all \cc{zval} objects allocated through this chain
share the same type
(measured by the two innermost return addresses
on the call stack).
This leaves the dangling pointer plenty of
candidate objects to refer to
after several rounds of deserialization in PHP.
TypeAfterType can inline \cc{malloc} wrappers
but the inlining stops at \cc{php\_var\_unserialize}
because it sees the \cc{sizeof(zval)} argument in \cc{emalloc}
and hence,
will allocate all \cc{zval} objects
originating from this \cc{malloc} wrapper from the same pool.
Unfortunately,
the dangling pointer is also allocated this way,
enabling UAF among the dangling pointer to other \cc{zval} objects as well.

\sys can mitigate this exploit because
\sema is not only context-sensitive but also flow-sensitive.
For examples,
a session initialization \cc{zval} can never be allocated from the same pool as a \cc{zval} created in the middle of a session.

\section{Performance Evaluation}
\label{s:eval}

We evaluate the performance of \sys across a diverse range of scenarios,
including macro, micro, and real-world programs.
For comparative analysis, we also benchmark 
two type-based allocators: 
FFMalloc~\cite{ffmalloc} and TypeAfterType~\cite{typeaftertype},
two complete UAF-mitigating allocators:
MarkUs~\cite{markus} and MineSweeper~\cite{minesweep},
and the \glibc memory allocator~\cite{glibc-malloc}
on the same test suites.
We additionally compare with 
DangZero~\cite{dangzero} on all benchmarks and
PUMM~\cite{pumm} on targeted server programs (\autoref{ss:eval_real})
as both are state-of-the-art
UAF-mitigating allocators with low overheads
despite DangZero requires kernel privilege and
PUMM requires annotation and profiling.
Although Cling~\cite{cling} is
a closely related work,
we omit it in our evaluation because
its code is not available.

Based on the design features of each memory allocator,
we expect that \sys should incur a:

\begin{enumerate}[noitemsep,topsep=0pt,leftmargin=10pt]
  \item[A] lower run-time overhead than MarkUs and MineSweeper,
  \item[B] lower memory overhead than FFMalloc,
  \item[C] similar run-time overhead compared with TypeAfterType
  and potentially a higher memory overhead,
  \item[D] smaller run-time overhead and similar memory overhead with DangZero,
  \item[E] similar run-time and memory overheads with PUMM.
\end{enumerate}

\PP{Preview}
The outcomes of our evaluation are in
alignment with these expectations
with consistent results across
mimalloc-bench,
SPEC CPU 2017,
PARSEC 3, and
real-world server programs (Nginx, Lighttpd, and Redis).

\subsection{Evaluation Setup}
\label{ss:eval_setup}
All experiments except DangZero are conducted 
in the Ubuntu 22.04.4 environment,
on a server configured with a 48-core
2.40GHz Intel Xeon Silver 4214R CPU with
128GB of system memory.
DangZero experiments are executed on the same machine
with QEMU-KVM
as DangZero requires patching the Linux 4.0 kernel in the guest VM.

We use LLVM 15
to compile programs.
For simplicity,
we use the WLLVM~\cite{wllvm} compiler wrapper
to link the whole program bitcode into a single IR file.
While generating the IR, 
we enable the compiler  
to track pointer types by setting the 
\cc{-fno-opaque-pointers} flag,
and disable constructor aliasing
with \cc{-mno-constructor-aliases} flag
to simplify the call graph.
We then transform the IR using our pass and 
compile it to 
generate the hardened program.
We also generate unhardened
programs by directly compiling the IR
without running the \sys-specific
transformation pass.

We use a Python wrapper
to measure clock time and
maximum memory usage (\cc{maxrss})
in program execution
for all test programs
except from DangZero-protected programs,
which is additionally measured with the page table size as instructed in their paper.
All results are based on five runs, 
normalized with respect to the corresponding \glibc.
We compute performance averages using geometric means,
and report standard deviations as well.

While we built all related tools as instructed
in the latest versions of their respective
official GitHub repositories,
we note that
MarkUs, TypeAfterType, DangZero, and MineSweeper are not compatible 
with all tests.
We exclude them from computing their respective average
overheads and the complete list can be found at~\autoref{s:app_fail}.
Individual test results are at~\autoref{app:individual_results}.
Maximum working set size is at~\autoref{app:wss}.

\subsection{Macro Benchmarks}
\label{ss:eval_macro}

\begin{figure}
    \centering
    \includegraphics[width=.45\textwidth]{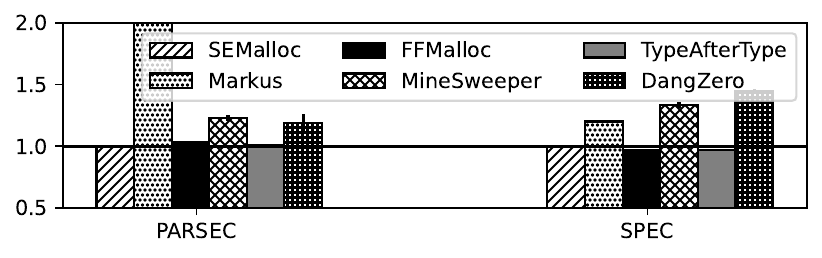}
    \caption{Normalized average and standard deviation of run-time overhead
        on PARSEC and SPEC benchmarks.}
    \label{fig:macro-t}
\end{figure}

\begin{figure}
    \centering
    \includegraphics[width=.45\textwidth]{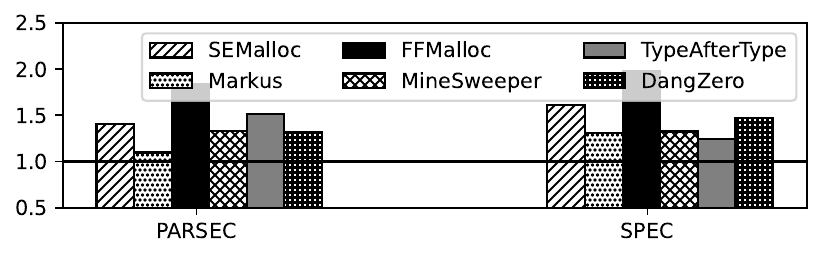}
    \caption{Normalized average and standard deviation of memory overhead
        on PARSEC and SPEC benchmarks.}
    \label{fig:macro-m}
\end{figure}

We choose the widely used SPEC and PARSEC
benchmark suites as macro benchmarks.
They are general-purpose benchmarks
with various kinds of programs
that can show the performance of \sys
in a broad range of scenarios.

\emph{SPEC CPU2017}:
We use SPEC CPU2017~\cite{spec} version 1.1.9 and
report the results of 12 C/C++ tests in
both "Integer" and "Floating Point" test suites.
We note that some tests run 
the executable multiple times with 
different inputs.
We report the sum of time and
the max of memory use for these tests.

\emph{PARSEC 3}:
We use the latest PARSEC 3~\cite{parsec} benchmark, 
excluding two (``raytrace'' and ``facesim'') from analysis because they are incompatible with the Clang compiler,
and one (``x264'') as it causes a segmentation fault
with \glibc.

\PP{Benchmark Performance}
We provide the run-time and memory overheads
as well as standard deviations
of these benchmarks 
in~\autoref{fig:macro-t} and~\autoref{fig:macro-m}.
Performance results of individual programs
are in~\autoref{tab:macro-individual-t} and~\autoref{tab:macro-individual-m} .

On SPEC,
\sys, FFMalloc, and TypeAfterType outperform
the \glibc allocator (0.6\%, 3.3\%, and 3.0\% respectively).
This is explainable as pre-allocating
heap pools by types reduces the number of page requests
made the kernel and hence can reduce allocation latency.
Placing heap objects of similar types or \sema{s}
in adjacent memory is also beneficial to cache lines.
MarkUs and MineSweeper incur significant overheads (21.0\% and 33.4\% respectively),
which is expected due to expensive pointer scanning operations.
DangZero also incurs a significant 45.2\% overhead even with a modified kernel presented,
which does not align with our expectations,
and is explained below.

All allocators incur extra memory overhead than \glibc.
As expected,
for type-based allocators,
the more sensitive the type
(TypeAfterType $\rightarrow$ \sys $\rightarrow$ FFMalloc)
the greater the memory overhead 
(23.5\% $\rightarrow$ 61.0\% $\rightarrow$ 98.4\%).
MarkUs and MineSweeper incur 31.1\% and 32.5\% memory overheads 
respectively due to
quarantine of freed blocks
although the number here is for reference only.
DangZero incurs a 47\% memory overhead
(including kernel memory consumption)
due to the use of alias page tables.

The results on PARSEC also align with expectations that \sys incurs:
smaller run-time overhead (-0.4\%) than MarkUs (144\%) and MineSweeper (23.0\%),
smaller memory overhead (40.5\%) than FFMalloc (84.1\%),
similar run-time overhead with TypeAfterType (1.0\%), and
smaller run-time and similar memory overheads with DangZero (19.5\% and 32.3\% respectively).

\PP{Abnormalities}
While the overall evaluation results
align with expectations A, B, C, and D,
we do notice abnormalities in the results.
Failed test cases and how they might affect
the reported evaluation numbers in related works
are summarized in~\autoref{s:app_fail}.
Here, we focus on discussing
individual test cases that do not
yield expected results.

\sys allows memory reuse among allocations of the same \sema
while FFMalloc does not allow any virtual memory reuse,
thus running \sys should 
incur less memory overheads compared with FFMalloc.
However, on the benchmarks,
we observe three exceptions: ``641'', ``644'', and ``fer''.
Test ``641'' and ``fer'' frequently
call functions in external libraries 
that allocates heap memories
causing excessive memory use.
Test ``644'' reaches its memory usage peak
at the beginning of the program
that allocates a significant 
number of blocks together and 
they are all not released
until the end of the program.
As FFMalloc allocates blocks at a 16-byte granularity,
it uses less memory to allocate them 
compared with \sys uses the size of two size classes
to allocate blocks.
The observed overhead comes from the data storage instead
of the way \sys reuse freed blocks.

\subsection{Micro Benchmarks}
\label{ss:eval_micro}

\begin{figure}[t] 
    \centering
    \includegraphics[width=.43\textwidth]{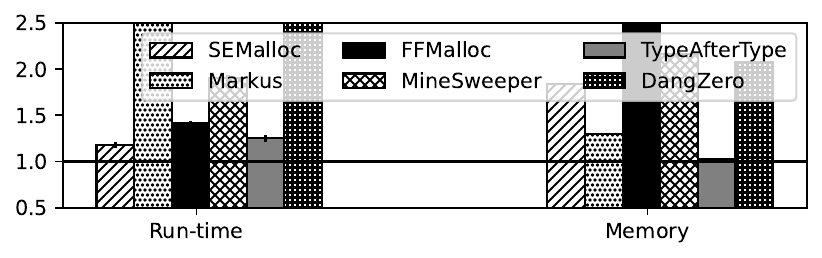}
    \caption{Normalized average and standard deviation of run-time and memory overheads on mimalloc-bench.}
    \label{fig:mimalloc}
\end{figure}

We use mimalloc-bench~\cite{mimalloc-bench},
a dedicated benchmark designed to stress test memory allocators
with frequent (and sometimes only)
allocations and de-allocations.
We exclude one test: ``mleak'' that tests 
memory leakage instead of allocation performance,
and summarize the overheads and standard deviations
of the rest of tests in~\autoref{fig:mimalloc}.
Individual results can be found in~\autoref{tab:micro-individual-t} and~\autoref{tab:micro-individual-m}.

\begin{figure}[t]
    \centering
    \includegraphics[width=.43\textwidth]{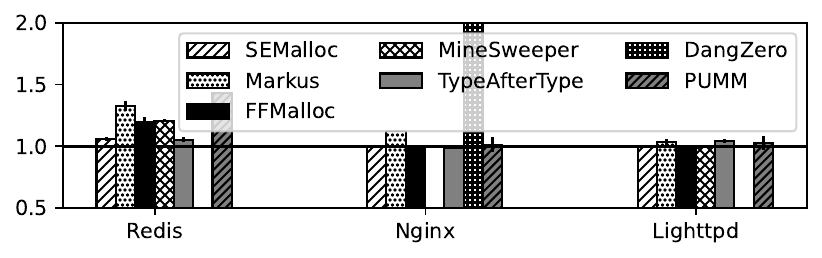}
    \caption{Normalized average and standard deviation of throughput overhead
        on three real-world programs.}
    \label{fig:realworld-t}
\end{figure}

\begin{figure}[t]
    \centering
    \includegraphics[width=.43\textwidth]{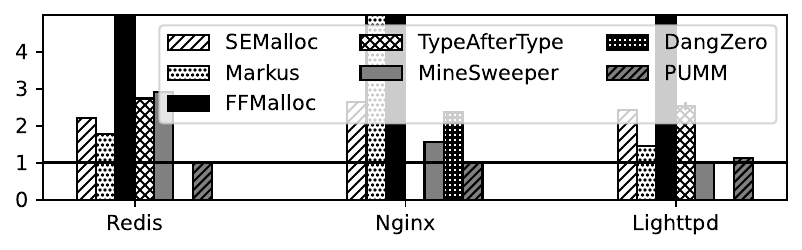}
    \caption{Normalized average and standard deviation of memory overhead
        on three real-world programs.}
    \label{fig:realworld-m}
\end{figure}

On average,
\sys introduces less execution delay
compared with allocators that offer more security
(i.e., MarkUs, MineSweeper, DangZero and FFMallloc) and
perform slightly better than TypeAfterType.
For memory overhead,
\sys cuts the memory usage by more than half 
compared with FFMalloc,
which aligns with our expectations and 
make it a possible approach for real-world programs.

\subsection{Performance on Real-world Programs}
\label{ss:eval_real}

We evaluate three real-world
performance of \sys using
Nginx (1.18.0), Lighttpd (1.4.71)
and Redis (7.2.1).
For network servers,
we use ApacheBench (ab)~\cite{ab} 2.3 to evaluate 
their throughput with 500 concurrent requests,
and take the Nginx default
613 bytes root page as the requested page.
On Redis, we use the same settings as how its performance
is measured in mimalloc-bench~\cite{mimalloc-bench}. 

The results are in~\autoref{fig:realworld-t} and~\autoref{fig:realworld-m}.
Overhead numbers can be found in~\autoref{tab:realworld-individual}.
While running Nginx,
MarkUs consumes a significant amount of memory
possibly due to an implementation error.
DangZero is not compatible with Redis and Lighttpd,
and MineSweeper is not compatible with Nginx.
Running them causes segmentation faults and
hence we exclude them from the analysis.
PUMM incurs
negligible run-time overheads for the two web servers but
an abnormal 43\% overhead for Redis,
possibly due to an implementation bug
or an incomplete program profiling 
that misidentifies the ``task''.
Albeit these outliers,
the results align with our expectations
for \sys set earlier in the beginning of~\autoref{s:eval}.

\section{On Recurrent Allocations}
\label{ss:eval-other-stats}

In \sys,
a \sema only needs to be tracked dynamically if
heap objects of this \sema are allocated recurrently,
i.e., through loops or recursions
(see~\autoref{ss:design_cyclic},
\autoref{ss:overview-repr}, and \autoref{ss:design_enforce}).
For non-recurrent allocations, 
once an object is freed,
its space is never reused.
In two extreme cases,

\begin{itemize}[noitemsep,topsep=0pt,leftmargin=*]
\item if a program itself involves absolutely zero
    recurrent heap allocations
    (but the dependent libraries may allocate heap memories)
    \sys behaves exactly like FFMalloc~\cite{ffmalloc};

\item if there is only one execution context where
    heap allocation can happen (i.e., a single \sema),
    \sys behaves exactly like the glibc
    heap allocator~\cite{glibc-malloc}.
\end{itemize}

\noindent
Fortunately,
most programs are not written in these extreme cases
as shown in the last two columns of~\autoref{tab:allocations}.
And yet,
this observation leads us to wonder
how prevalent recurrent heap allocations are in
common benchmark programs that evaluate heap allocators.
Needless to say,
programs that have a more diverse set of recurrent allocations
can benefit more from the fact that
\sys attempts to strike a sweet spot in
security, performance, and memory overhead
in the context of UAF mitigation.

We use recurrent allocation percentage 
to describe how many allocations are one-time allocations.
For most programs that frequently allocate blocks, 
over 99\% of the allocations are effectively captured
and allocated to individual \sema pools.
These pools handle a significant amount of
memory reallocation (as shown in the fourth to the last column),
which improves memory efficiency and thus
explains why empirically \sys
incurs a lower memory overhead than FFMalloc
and limits memory leakage.

However,
we observe three exceptions: "620," "bod," and "fer".
They often call functions from external libraries
(such as those linked with the \cc{-lm} flag in the \cc{math.h} library)
that allocate heap memory as well.
These external libraries are not transformed by \sys,
leading to untracked heap allocations that are handled
in the global non-releasing pool (like FFMalloc).
Therefore,
adopting \sys for a
program that heavily depends on external libraries
for heap allocations may not be ideal,
and the developers can opt to recompile the dependent libraries
with \sys for better compatibility.

\section{Concluding Remarks}
\label{s:conclusion}

Type is a loosely defined concept
in security research and
is often subject to different interpretations.
In this paper,
we look at ``type'' through the lens of heap allocators.
To a heap allocator, a type is an encoding of information
about the to-be-allocated object.
Intuitively,
the more information (i.e., semantics)
a heap allocator knows,
the better decisions it can make.
While conventional memory allocators
only take object size as the semantics,
we argue that \sema can be a useful extension
to object size, and,
more importantly,
can be deduced cheaply at runtime\footnote{Memory overhead
of \sys is not related to deducing \sema.}.

Through \sys,
we show that \sema can be used to balance
security, run-time cost, and memory overhead
in UAF mitigation.
And yet we believe that
\sema is applicable beyond \sys.
For example,
when applied to a performance-oriented
memory allocator,
\sema might help the allocator to
partition heap pools strategically
to exploit cache coherence.
Alternatively,
combined with other
software fault isolation (SFI)
strategies (e.g., red-zoning or
pointer-as-capabilities),
\sema might help specify and enforce
finer-grained data access policies.

\PP{Acknowledgement.} 
This work is funded in part by NSERC (RGPIN-2022-03325), the David R. Cheriton endowment, and research gifts from Intel Labs.

\bibliographystyle{ACM-Reference-Format}
\bibliography{p,sslab,conf}

\appendix
\section{Appendix}

\subsection{Code Example}
\label{app:code_example}

This section presents a hypothetical code snippet
that yields the call graph presented in~\autoref{fig:example}.

\begin{figure}[h]
    \centering
    \small
    \hfill\begin{minipage}[c]{.925\linewidth}\input{code/code-example}\end{minipage}
    \caption{Code snippet that yields the call graph in~\autoref{fig:example}.}
    \label{fig:code-example}
\end{figure}

\subsection{Instruction Insertion Summary}
\label{app:insertion}

The number of LLVM IR instructions instrumented
at different code locations
are summarized in~\autoref{tab:trans}.

\begin{table}[h]
    \centering
    \small
    \begin{tabular}{lrr}
    \toprule
        \textbf{Code location} & \textbf{Call} & \textbf{Invoke} \\\midrule
        SCC inbound edges & 1 & 2 \\
        SCC inner edges & 9 & 13\\
        SCC outbound edges & 16 & 22\\ \midrule
        Iterative node & 6 & 12\\
        Branch node & 6 & 20\\
        \cc{malloc} call site & 12 & 20\\\midrule
        Duplicated invoke node & \multicolumn{2}{c}{2*calls in the same bracket} \\
        \bottomrule
        
    \end{tabular}
    \caption{Number of instructions inserted for \cc{call}, \cc{invoke}, and for duplicating the invoke nodes.
    \textnormal{In the call graph,
we use ``branch node'' to denote a node with
more than one incoming edges and
``iterative node'' to denote a node that has
at least one outgoing edge annotated in dashes
(i.e., the call site is in a loop).
We note that a branch node can potentially
also be an iterative node.
In this case, both groups of instructions will be inserted.
}}
    \label{tab:trans}
\end{table}

Briefly,
following is a summary of instructions added:

\begin{itemize}[noitemsep,topsep=0pt,leftmargin=*]
\item For an SCC inbound edge,
instructions are inserted after
the call site to clear $s$.

\item For an intra-SCC edge,
instructions are inserted before 
and after the call site to update $s$.

\item For an SCC outbound edge
instructions are inserted before the call
to compute $h$ (which is \rid)
and clear $s$.

\item For an iterative node,
instructions are inserted before
and after the call to maintain \nid.

\item For a branch node,
instructions are inserted
before and after the call to maintain \nid.

\item For a \cc{malloc} call site,
instructions are inserted before the call to
encode \rid and \nid into the size parameter.
\end{itemize}
\vspace{2pt}

\noindent
Additionally,
if a function is called
with exception handling
(via the \cc{invoke} instruction in LLVM),
additional instructions need to be inserted
to handle the unwind branch and to duplicate 
the execution logic to make it compatible with \sys.
We refer the readers to~\autoref{app:invoke}
for details.

\subsection{Formal Analysis}
\label{app:sec_formal}

\sys stops UAF attacks by defining the object types 
based on the path used to allocate them,
and guarantees that 
upon the allocation call site (i.e., \cc{malloc} call),
the executing tracing and weight assignment can
uniquely identify each type.
As explained in~\autoref{ss:design_identify},
\sema is classified into two types in \sys:
allocated through recurrent allocation trace (RA), 
and non-recurrent regular allocation trace (NA).
There are three possibilities if \sys
cannot represent each \sema identically
based on this classification:
\begin{enumerate}
    \item NA1 and NA2;
    \item NA1 and RA1;
    \item RA1 and RA2.
\end{enumerate}

Next, we explain why \sys
can notify the runtime wrapper
to allocate blocks of different \sema separately 
in each of the above scenario.

\subsubsection*{NA1 and NA2}
A \sema is classified as 
NA if there is no recurrent call
site in its allocation trace.
Thus, there is no possibility that 
a NA trace accidentally classified as RA.
NA blocks are not to be reused by another block
after freed.
As a result, the two NA blocks are never going 
to share the same memory address.

\subsubsection*{NA1 and RA1}
As explained above,
a NA trace is not classified as RA 
thus never reuses a block allocated 
to others or releases itself
to be used by another object.
Thus, cross-category memory address reuse
also will not happen.

\subsubsection*{RA1 and RA2}
RA blocks release the memory space after being freed
and a further block with the same \sema will 
be allocated to this address.
If two different RA objects allocated to the same virtual 
memory address,
their \nid must be the same (i.e., $t_1 = t_2$)
upon calling \cc{malloc}.
We define the call trace of RA1 as 
$C1 = \alpha_1, \alpha_2, ..., \alpha_n$,
and the call trace of RA2 as
$C2 = \beta_1, \beta_2, ..., \beta_m$.
Our target is to prove:

\PP{Theorem} Given any two recurrent call traces
$C1$, $C2$, where $C1 \neq C2$. 
Upon the \cc{malloc}
call site $\beta_m$ of $C2$, its \nid $t_2$
is never the same to the \nid of $C1$ at their
\cc{malloc} call site $\alpha_n$ with the value $t_1$.

To prove this, we first need to prove:

\PP{Lemma 1} Given any recurrent call trace
$C = \gamma_1, \gamma_2, ..., \gamma_r$,
its \nid $t$ is always the same at its \cc{malloc} call site $\gamma_r$.

Suppose the iterative call site is $\gamma_i$.
After returns $\gamma_r$ and before calling it the next time,
function call 
$\gamma_r, \gamma_{r - 1}, ..., \gamma_i$ sequentially returns and
$\gamma_i, \gamma_{i + 1}, ..., \gamma_r$ are sequentially called.
As $t$ is increased with a specific value before a call site
and is decreased with the same value after the call returns,
executing the above sequence does not change the $t$.
Thus, we have proved the lemma 1.

\PP{Lemma 2} The weight of function $w_f$ is larger than 
the sum of the weights on any trace starting with it.

We can use mathematical induction to simply prove this.
Suppose $f$ does not call any functions other than \cc{malloc},
its weight is the number of calls to \cc{malloc} it has,
and the path weights are assigned from zero to $w_f - 1$.
Now, suppose $f$ does call functions other than \cc{malloc},
we suppose that $f$ calls $F = {f_1, f_2, ..., f_q}$ sequentially.
We assume that $\forall \Tilde{f} \in F$, lemma 2 applies.
We take an arbitrary $f_p \in F$.
The weight assigned with it is $\sum_{i=1}^{p-1}wf_{i}$,
and by assumption the maximum path weight among all call traces within $f_p$ is 
$\sum_{i=1}^{p}wf_{i} < wf = \sum_{i=1}^{q}wf_{i}$
Thus, we have proved the lemma 2.

\PP{Prove the theorem}
To prove the theorem, We only need to show that the sum of the weights
in each trace is identical.
We denote the weight of $\alpha_i$ as $w\alpha_i$, and
the weight of $\beta_i$ as $w\beta_i$.
We assume that the two traces shares the first $k$ prefixes
(i.e., $\alpha_i = \beta_i, 0 < i \leq k$).
Now take the $(i+1)$th call of each trace $\alpha_{i+1}$ and $\beta_{i+1}$,
their caller is the same function.
Without the lose of generality,
we suppose that $w\alpha_{i+1} > w\beta_{i+1}$.
According to the weight allocation algorithm~\autoref{alg:weight_assign},
in the caller function,
$\beta_{i+1}$ is called before $\alpha_{i+1}$,
and $w\alpha_{i+1} - w\beta_{i+1} \geq w_{\delta}$,
where $w_{\delta}$ is the weight of the function
called by $\beta_{i+1}$.
According to lemma 2, any function traces within $\delta$
is smaller than $w_\delta$,
thus $w\alpha_{i+1} > \sum_{k=i+1}^{m}w\beta_{k}$.
Thus, $\sum_{k=1}^{n}w\alpha_{k} > \sum_{k=1}^{m}w\beta_{k}$.
We have proved the theorem.

\subsection{List of Supported Allocation APIs}
\label{app:support_api}

\sys supports the following allocation APIs:
\cc{malloc},
\cc{calloc},
\cc{realloc},
\cc{memalign},
\cc{pthread\_memalign}, and
\cc{aligned\_alloc}.

\subsection{Transformation for Function Call
with Exception Handling}
\label{app:invoke}
In LLVM,
regular function calls are represented with the
\cc{call} instruction.
This instruction is similar to a regular function call
in high-level programming languages
and does not encode exception handling semantics.
For calls that may throw an exception,
LLVM uses the \cc{invoke} instruction.

Different from the \cc{call} instruction that 
returns the control flow to the next instruction,
\cc{invoke} terminates the control flow
and jumps to two destinations that 
contains the regular branch and the 
exception handling branch (a.k.a., the unwind branch).
If more than one function calls are made
in one exception-handling context 
(e.g., more than one functions calls
in the same \cc{try} block in C++),
there is still only one unwind branch
that all \cc{invoke} instructions will jump to.

When making a regular call,
\nid is decreased after the \cc{call} returns.
With the \cc{invoke} instruction,
\nid needs to be decreased in both
destination branches.
The unwind branch also needs to be exclusive
to each \cc{invoke} as \nid needs to be reduced 
with a different value in different sites.
To achieve this,
we duplicate the unwind branch and 
guarantee that each branch is only
jumped from one \cc{invoke} instruction.

The unwind branch might contain $\phi$-instructions,
whose return value is dependent to the prior 
basic block the control flow jumped from.
To make the transformation
compatible with this special  instruction,
we only duplicate the basic exception handling logic
(first half of the unwind basic block) and 
insert instructions to reduce \nid
for the unwind branches here.
We create a new basic block only contains
the second half of each branch, and all
$\phi$-instructions are in the newly created basic block.

Similarly, 
a \cc{invoke} destination block can
have incoming edges from basic blocks
that do not end with the \cc{invoke} instruction.
We need to duplicate this destination block
as well to avoid always
executing the inserted tracking instructions even
this basic block is not jumped from a \cc{invoke} call site.
We create a new basic block and insert the \sema tracking 
instructions here.
We then replace the invoke destination to this block and
link this block with the old destination,
while update all $\phi$-instructions accordingly.

\subsection{Additional Exploitation Case Studies}
\label{app:additional_cve}

\PP{Python-24613}~\cite{python-24613}
This vulnerability resides 
in the logic of parsing an array from a string and appending it to an existing array.
In the \cc{array.fromstring()} method,
the Python interpreter calls \cc{realloc} to guarantee that
the allocated memory of the appended array object is big enough,
and calls \cc{memcpy} to copy the data 
from the string to the new array.
However, if the array is appending itself,
i.e, the string and the array is the same object,
\cc{realloc} essentially frees this object,
and the subsequent \cc{memcpy}
copies the freed heap chunk to the new object.
An attacker can exploit this vulnerability
by racing to allocate objects filled with attacker-controlled
malicious data over the freed chunk,
making this a type-C exploit (see~\autoref{ss:bg-exploit}).

In this exploit, both the dangling pointer and the target object (i.e., the object that the attacker uses the dangling pointer to read from) are allocated through
\cc{array\_fromstring}$\rightarrow$\cc{realloc}.
\cc{array\_fromstring} is an exposed Python API that can be called directly in a Python script,
and is only called by one other function, \cc{array\_new},
which is also a Python API.
All three allocators here can differentiate these two call sequences,
thus providing complete protection for this vulnerability.

\PP{CVE-2012-0469}~\cite{typeaftertype}
This vulnerability is in the \cc{indexedDB} module of Firefox.
While a \cc{IDBKeyRange} is freed,
its reference is left in the object table.
The attacker can craft an object,
for example, a \cc{vector} to reclaim the pointed space,
and interpreting the crafted object using the dangling pointer
can cause arbitrary code execution. 
This is a type-B exploit.

As this object can only be allocated by calling \cc{new} to its constructor,
Cling can stop the type confusion on the primary type,
provides a partial protection.
\sys and TypeAfterType can further
differentiate the source of the created \cc{IDBKeyRange} object,
from serialized data or explicitly created in the JavaScript script,
thus providing complete protection for this vulnerability.

\subsection{Implementation Details of The Allocation Backend
in \NoCaseChange{\sys}}
\label{app:backend}

The heap allocator backend of \sys
is implemented using the 
\cc{dlsym} function and is a dynamic library
that can be loaded by either setting the \cc{LD\_PRELOAD}
environment variable to replace the system default memory allocator
or direct linkage during compilation.

\sys maintains a per-thread metadata that stores 
the status of all blocks allocated in this thread.
If a block not allocated in this thread is freed,
\sys will atomically add this block to the free-list
of the thread that allocates it and defers the deallocation
until that thread handles a heap memory management operation.
Other than the free-list, for each thread
\sys maintains a global pool for all one-time allocations,
a lazy pool for all first-seen recurrent allocations,
several individual pools for each recurrent allocations,
and a map that used to locate the pool for each \sema.
The lazy pool and global pool are pools for all size classes,
while individual pools contain a limited set of size classes
(most of the time only one)
that have been used to save virtual memory address space.

Upon creation,
each pool is allocated with a dedicated virtual memory
address range that all its memory will be allocated from.
For each sub pool of the global pool or the lazy pool,
\sys allocates the block sequentially.
For each individual pool,
\sys maintains a free list as well and will allocate 
the head of the free list if it is not empty.
Otherwise,
\sys will also allocate the mapped memory of this individual pool
sequentially.

For each block,
a 16-byte metadata is stored immediately before the data.
For huge block, it stores the block type (huge) and the block size.
For regular small blocks, 
it stores the block type (regular) with one byte,
the ID of the thread that allocates the block with two bytes,
the pointer to the pool that allocates the block with eight bytes,
and an offset if the block is allocated via \cc{memalign}-like functions
to locate the start byte of the block chunk with four bytes.

When a \cc{malloc} call comes,
\sys firstly check if the block size
is larger than the huge allocation threshold or not.
If so,
\sys uses \cc{mmap} to allocate this block and sets the header.
Otherwise, \sys takes the recurrent identifier.
If it is not set, 
\sys will use the global pool to allocate this block.
Otherwise,
\sys will take \sema and check if a block with it
is already allocated or not using the lazy pool.
If so,
\sys can confirm that this blocks with this \sema is recurrently allocated
and will
allocate an individual pool for all following allocations with this \sema.
The map will be updated with this new entry as well.
Otherwise,
the lazy pool will allocate this block.

When a \cc{free} call comes,
\sys will firstly take the header of the pointed block
to check if it is a huge block or not
If so,
\sys uses \cc{munmap} to deallocate this block.
Otherwise,
\sys will check the thread id and put it to the corresponding free list 
if this block is not allocated in this thread.
If the block is allocated in this thread,
\sys then takes the pool pointer.
If the pool is a global pool or lazy pool,
\sys will release the taken memory to the operation system
by calling \cc{madvise}.
However,
this virtual memory address will not going to allocated to any blocks again.
If the pool is an individual pool,
\sys will put this block to the pool's free list,
recycling it for another block with the same \sema.

\PP{Memory leakage}
In~\autoref{tab:allocations},
we use the third to the last column to list
memory leakage caused by \sys.
We show that programs
not using memory-allocating external libraries
have negligible memory leakage.

\subsection{List of Failed Tests and Influences on the Average Overheads}
\label{s:app_fail}

\PP{PARSEC}(referenced by the first three letters of its name)
\begin{itemize}
    \item MarkUs:
    \begin{itemize}[noitemsep,topsep=0pt,leftmargin=*]
        \item fer (\cc{SIGSEGV})
    \end{itemize}

    \item MineSweeper:
    \begin{itemize}[noitemsep,topsep=0pt,leftmargin=*]
        \item fer (\cc{SIGSEGV})
    \end{itemize}
    
    \item TypeAfterType:
    \begin{itemize}[noitemsep,topsep=0pt,leftmargin=*]
        \item can (incomplete type tracking support)
        \item fer (LLVM pass exception)
        \item swa (incomplete type tracking support)
        \item vip (incomplete type tracking support)
    \end{itemize}

    \item DangZero:
    \begin{itemize}[noitemsep,topsep=0pt,leftmargin=*]
        \item bod (\cc{SIGSEGV})
        \item ded (\cc{SIGSEGV})
        \item fer (\cc{SIGSEGV})
        \item vip (assertion fail)
    \end{itemize}
\end{itemize}

\PP{SPEC}
\begin{itemize}
    \item TypeAfterType:
        \begin{itemize}[noitemsep,topsep=0pt,leftmargin=*]
        \item 602 (LLVM pass exception)
    \end{itemize}

    \item DangZero:
    \begin{itemize}[noitemsep,topsep=0pt,leftmargin=*]
        \item 638 (\cc{SIGSEGV})
        \item 644 (\cc{SIGSEGV})
        \item 657 (\cc{SIGSEGV})
    \end{itemize}
\end{itemize}

\PP{Mimalloc-Bench}
\begin{itemize}
    \item TypeAfterType:
    \begin{itemize}[noitemsep,topsep=0pt,leftmargin=*]    
        \item alloc-test (incomplete type tracking support)
        \item malloc-large(incomplete type tracking support)
        \item rbstress (configuration failure)
    \end{itemize}
    \item DangZero:
    \begin{itemize}[noitemsep,topsep=0pt,leftmargin=*]    
        \item rbstress (\cc{SIGSEGV})
        \item sh6bench (\cc{SIGSEGV})
        \item sh8bench (\cc{SIGSEGV})
    \end{itemize}
\end{itemize}

\PP{Real-world programs}
\begin{itemize}
    \item MineSweeper:
    \begin{itemize}[noitemsep,topsep=0pt,leftmargin=*]    
        \item Nginx (\cc{SIGSEGV})
    \end{itemize}
    \item DangZero:
    \begin{itemize}[noitemsep,topsep=0pt,leftmargin=*]    
        \item Redis (\cc{SIGSEGV})
        \item Lighttpd (\cc{SIGSEGV})
    \end{itemize}
\end{itemize}

MarkUs and MineSweeper
fail to run on one PARSEC test (``fer'').
We exclude it
from computing the average overheads for
MarkUs and MineSweeper.
However,
we should expect smaller overheads
for them given that this failed test does not allocate
blocks frequently and running it with other allocators
incurs smaller overheads
Similarly,
for TypeAfterType, 
compiling ``vips'', ``swa'', ``can'', ``fer'',
and ``602'' fails
due to its incompatibility with the
\cc{using} keyword in C++ and 
incomplete support for variable type casts.
Running it with ``bod'' uses more than ten times of memory
than the baseline,
potentially due to an implementation bug.

Additionally, running
``bod'', ``ded'', ``fer'', ``638'', ``644'', and ``657'' with DangZero fails
due to segmentation faults and 
``vip'' fails due to an assertion failure.
This incompatibility causes the observed overhead to be larger than
those reported in the paper.
However, we note that the per-test numbers are close to their reported numbers
(see~\autoref{tab:macro-individual-t} and~\autoref{tab:macro-individual-m}).
The failed SPEC tests all have small overheads that cause the average overheads to become larger.

\subsection{Run-time and Memory Overhead Details of
Each Benchmark Test}
\label{app:individual_results}

\PP{Macro benchmark}
SPEC and PARSEC in~\autoref{tab:macro-individual-t} and~\autoref{tab:macro-individual-m}.

\PP{Micro benchmark}
mimalloc-bench in~\autoref{tab:micro-individual-t} and~\autoref{tab:micro-individual-m}.

\PP{Real-world programs}
Three real-world programs
(Redis, Nginx, Lighttpd)
in~\autoref{tab:realworld-individual}.

\begin{table*}[h]
\small
\centering
\begin{tabular}{l|cccccc}
\toprule
Test ID & SEMalloc & Markus & FFMalloc & MineSweeper & TypeAfterType & DangZero \\
\midrule
600 & 1.14 (0.00) & 1.08 (0.01) & 1.02 (0.01) & 1.14 (0.06) & 1.01 (0.01) & 1.47 (0.12) \\
602 & 1.01 (0.00) & 1.05 (0.01) & 0.58 (0.00) & 3.07 (0.06) & -           & 1.70 (0.01) \\
605 & 0.97 (0.00) & 1.01 (0.00) & 0.99 (0.00) & 1.02 (0.07) & 1.00 (0.00) & 1.11 (0.01) \\
619 & 1.00 (0.01) & 1.01 (0.01) & 1.01 (0.01) & 1.06 (0.12) & 1.21 (0.01) & 1.02 (0.00) \\
620 & 1.01 (0.01) & 2.20 (0.04) & 1.04 (0.01) & 1.70 (0.10) & 1.03 (0.01) & 2.30 (0.00) \\
623 & 0.84 (0.00) & 2.38 (0.04) & 1.03 (0.00) & 2.79 (0.08) & 1.01 (0.01) & 3.01 (0.05) \\
625 & 1.00 (0.00) & 1.01 (0.00) & 1.01 (0.00) & 1.01 (0.00) & 1.01 (0.00) & 1.11 (0.04) \\
631 & 0.99 (0.00) & 1.02 (0.00) & 1.00 (0.00) & 1.00 (0.00) & 1.01 (0.00) & 1.06 (0.00) \\
638 & 0.98 (0.00) & 1.00 (0.00) & 1.00 (0.00) & 1.00 (0.00) & 1.00 (0.03) & -           \\
641 & 1.03 (0.00) & 1.36 (0.01) & 1.04 (0.00) & 1.35 (0.01) & 1.02 (0.00) & 1.25 (0.02) \\
644 & 1.00 (0.00) & 1.15 (0.17) & 1.00 (0.00) & 1.36 (0.00) & 0.92 (0.01) & -           \\
657 & 0.98 (0.00) & 0.99 (0.00) & 0.99 (0.00) & 0.98 (0.01) & 1.01 (0.00)           & -           \\\midrule
Avg & 0.99 (0.00) & 1.21 (0.01) & \textbf{0.97 (0.00)} & 1.33 (0.02) & \underline{0.97 (0.00)} & 1.45 (0.02) \\\midrule
bla & 1.00 (0.01) & 1.00 (0.01) & 1.00 (0.01) & 0.98 (0.00) & 1.00 (0.01) & 1.04 (0.01) \\
bod & 1.00 (0.01) & 1.01 (0.01) & 1.00 (0.01) & 1.05 (0.00) & 1.00 (0.00) & -           \\
can & 0.99 (0.00) & 1.04 (0.01) & 0.99 (0.01) & 1.42 (0.02) &  -          & 1.33 (0.10) \\
ded & 0.98 (0.00) & 1412.44 (102.86) & 1.24 (0.00) & 1.73 (0.02) & 1.02 (0.02) &      - \\
fer & 1.01 (0.00) &  -          & 1.00 (0.00) &  -          &  -          & -           \\
flu & 1.00 (0.01) & 0.99 (0.00) & 1.00 (0.00) & 0.99 (0.00) & 1.00 (0.00) & 1.06 (0.01) \\
fre & 1.03 (0.01) & 1.07 (0.01) & 1.01 (0.01) & 1.02 (0.00) & 1.06 (0.00) & 1.44 (0.59) \\
str & 1.01 (0.00) & 1.02 (0.01) & 1.02 (0.01) & 1.16 (0.13) & 1.00 (0.00) & 1.10 (0.03) \\
swa & 1.00 (0.01) & 1.66 (0.02) & 1.06 (0.01) & 1.80 (0.02) &  -          & 1.29 (0.01) \\
vip & 1.00 (0.00) &  1.18 (0.00)         & 1.08 (0.00) &  1.19 (0.01)          &  -          &  -          \\\midrule
Avg & \textbf{1.00 (0.00)} & 2.44 (0.02) & 1.04 (0.00) & 1.23 (0.02) & \underline{1.01 (0.00)} & 1.19 (0.07) \\

\bottomrule
\end{tabular}
\caption{Normalized average run-time overheads (and standard deviations) of SeMalloc on SPEC and PARSEC. We indicate the best scheme in bold and the second best underlined to
show how SeMalloc hits the sweet spot in the tradeoff between run time and memory use.}
\label{tab:macro-individual-t}
\end{table*}

\begin{table*}[h]
\small
\centering
\begin{tabular}{l|cccccc}
\toprule
Test ID & SEMalloc & Markus & FFMalloc & MineSweeper & TypeAfterType & DangZero \\
\midrule
600 & 2.69 (0.00) & 1.65 (0.00) & 3.98 (0.00) & 1.74 (0.01) & 1.30 (0.00) & 1.85 (0.00) \\
602 & 1.08 (0.00) & 1.06 (0.00) & 1.27 (0.02) & 1.61 (0.02) & -           & 1.32 (0.00) \\
605 & 1.00 (0.00) & 1.28 (0.00) & 1.02 (0.00) & 1.01 (0.00) & 1.00 (0.00) & 1.25 (0.00) \\
619 & 1.00 (0.00) & 1.00 (0.00) & 1.02 (0.00) & 1.00 (0.00) & 1.00 (0.00) & 1.00 (0.00) \\
620 & 3.27 (0.00) & 1.98 (0.00) & 18.04 (0.00) & 2.17 (0.17) & 1.28 (0.00) & 3.32 (0.00) \\
623 & 1.58 (0.00) & 2.23 (0.00) & 3.54 (0.00) & 1.77 (0.02) & 1.00 (0.00) & 1.50 (0.00) \\
625 & 1.07 (0.00) & 1.12 (0.00) & 1.53 (0.00) & 1.08 (0.00) & 1.01 (0.00) & 1.09 (0.00) \\
631 & 1.00 (0.00) & 1.00 (0.00) & 1.01 (0.00) & 1.00 (0.00) & 1.00 (0.00) & 1.00 (0.00) \\
638 & 1.00 (0.00) & 1.07 (0.00) & 1.01 (0.00) & 1.06 (0.00) & 1.03 (0.02) & -           \\
641 & 4.29 (0.00) & 1.23 (0.01) & 3.98 (0.00) & 1.65 (0.05) & 1.89 (0.00) & 1.92 (0.00) \\
644 & 4.46 (0.00) & 1.76 (0.00) & 1.76 (0.00) & 1.42 (0.00) & 2.18 (0.00) & -           \\
657 & 1.00 (0.00) & 1.00 (0.00) & 1.02 (0.00) & 1.00 (0.00) & 1.01 (0.00)           & -           \\\midrule
Avg & 1.61 (0.00) & \underline{1.31 (0.00)} & 1.98 (0.00) & 1.33 (0.01) & \textbf{1.24 (0.00)} & 1.47 (0.00) \\\midrule
bla & 1.02 (0.00) & 1.00 (0.00) & 1.12 (0.00) & 1.01 (0.00) & 1.00 (0.00) & 1.00 (0.00) \\
bod & 2.50 (0.00) & 1.23 (0.01) & 3.14 (0.03) & 1.76 (0.01) & 10.21 (0.00)& -           \\
can & 1.21 (0.00) & 1.35 (0.00) & 1.23 (0.00) & 1.43 (0.00) &  -          & 1.24 (0.00) \\
ded & 1.05 (0.00) & 1.07 (0.00) & 1.05 (0.00) & 1.99 (0.00) & 1.07 (0.00) & -           \\
fer & 1.81 (0.00) & -           & 1.76 (0.00) & -           & -           & -           \\
flu & 1.04 (0.00) & 1.00 (0.00) & 1.13 (0.00) & 1.02 (0.00) & 1.00 (0.00) & 1.01 (0.00) \\
fre & 0.95 (0.00) & 0.99 (0.00) & 1.05 (0.00) & 0.96 (0.00) & 1.09 (0.00) & 1.02 (0.00) \\
str & 1.47 (0.00) & 1.09 (0.00) & 1.65 (0.00) & 1.08 (0.00) & 1.00 (0.00) & 1.07 (0.00) \\
swa & 2.37 (0.00) & 1.08 (0.00) & 9.77 (0.01) & 1.87 (0.07) &  -          & 3.83 (0.00) \\
vip & 1.47 (0.00) &  1.08 (0.00)          & 2.92 (0.00) &  1.53 (0.03)          &  -          &  -          \\\midrule
Avg & 1.41 (0.00) & \textbf{1.10 (0.00)} & 1.84 (0.00) & 1.33 (0.01) & 1.52 (0.00) & \underline{1.32 (0.00)} \\

\bottomrule
\end{tabular}
\caption{Normalized average memory overheads (and standard deviations) of SeMalloc on SPEC and PARSEC. We indicate the best scheme in bold and the second best underlined to
show how SeMalloc hits the sweet spot in the tradeoff between run time and memory use.}
\label{tab:macro-individual-m}
\end{table*}

\begin{table*}[h]
\small
\centering
\begin{tabular}{l|cccccc}
\toprule
Test ID & SEMalloc & Markus & FFMalloc & MineSweeper & TypeAfterType & DangZero \\
\midrule
alloc-test* & $1.20 (0.02)$ & $2.83 (0.08)$ & $1.51 (0.02)$ & $1.53 (0.04)$& - & 7.13 (0.14) \\
cscratch & $1.01 (0.00)$ & $1.00 (0.00)$ & $1.03 (0.00)$ & $1.05 (0.02)$ & 1.00 (0.00) & $1.62 (0.05)$ \\
cthrash & $1.01 (0.00)$ & $1.01 (0.00)$ & $1.03 (0.00)$ & $1.03 (0.01)$& 1.00 (0.00) & $1.75 (0.05)$ \\
glibc-simple & $1.16 (0.00)$ & $3.41 (0.04)$ & $1.29 (0.00)$ & $1.70 (0.07)$& 1.33 (0.01) & $16.39 (0.12)$ \\
malloc-large & $3.28 (0.00)$ & $3.96 (0.01)$ & $3.25 (0.01)$ & $3.19 (0.01)$ & - & $2.62 (0.02)$ \\
rptest* & $0.84 (0.13)$ & $8.43 (0.07)$ & $3.46 (0.01)$ & $9.94 (0.12)$& 1.23 (0.00) & $4.31 (0.05)$ \\
mstress & $1.28 (0.01)$ & $2.94 (0.02)$ & $2.53 (0.02)$ & $3.63 (0.09)$& 1.32 (0.23) & $3.90 (0.02)$ \\
rbstress & $1.03 (0.00)$ & $1.03 (0.00)$ & $1.02 (0.00)$ & $1.04 (0.01)$ & - & - \\
sh6bench & $0.95 (0.00)$ & $7.73 (0.13)$ & $1.51 (0.00)$ & $2.08 (0.05)$&  1.36 (0.00) & - \\
sh8bench & $0.89 (0.00)$ & $0.00 (0.00)$ & $1.60 (0.02)$ & $3.50 (0.13)$&  1.36 (0.03)& - \\
xmalloc-test* & $1.40 (0.04)$ & $1.33 (0.02)$ & $0.33 (0.02)$ & $0.47 (0.06)$ & 1.49 (0.03) & $4.50 (0.11)$ \\\midrule
Avg & \textbf{1.18 (0.03)} & 2.52 (0.04) & \underline{1.42 (0.02)} & 1.90 (0.04) & 1.25 (0.04) & 4.00 (0.06)\\
\bottomrule
\end{tabular}

\caption{Normalized average run-time overheads (and standard deviations) of \sys on mimalloc-bench (results of * marked tests use built-in measurements). We indicate the best scheme in \textbf{bold} and the second best \underline{underlined} to show how \sys hits the sweet spot in the tradeoff between run time and memory use.}
\label{tab:micro-individual-t}
\end{table*}

\begin{table*}[h]
\small
\centering
\begin{tabular}{l|cccccc}
\toprule
Test ID & SEMalloc & Markus & FFMalloc & MineSweeper & TypeAfterType & DangZero \\
\midrule
alloc-test & $2.72 (0.00)$ & $1.43 (0.00)$ & $58.99 (0.00)$ & $2.52 (0.05)$ & - & $13.45 (0.01)$ \\
cscratch & $1.99 (0.01)$ & $1.00 (0.01)$ & $8.04 (0.01)$ & $1.32 (0.01)$ & 1.01 (0.01) & $1.01 (0.01)$ \\
cthrash & $1.98 (0.01)$ & $1.00 (0.01)$ & $7.97 (0.01)$ & $1.30 (0.01)$ & 1.01 (0.01) & $1.02 (0.01)$ \\
glibc-simple & $3.67 (0.01)$ & $1.01 (0.01)$ & $8.02 (0.00)$ & $2.52 (0.08)$ & 1.01 (0.01)&  $3.46 (0.01)$ \\
malloc-large & $0.78 (0.00)$ & $0.78 (0.00)$ & $0.91 (0.00)$ & $0.80 (0.00)$ & - & $1.00 (0.00)$ \\
rptest & $3.56 (0.00)$ & $10.09 (0.01)$ & $8.28 (0.01)$ & $1.89 (0.01)$ & 1.28 (0.01)& $1.01 (0.01)$ \\
mstress & $1.43 (0.01)$ & $1.00 (0.01)$ & $8.04 (0.00)$ & $1.89 (0.05)$ & 1.00 (0.00)& $1.03 (0.00)$ \\
rbstress & $1.27 (0.00)$ & $1.09 (0.00)$ & $1.71 (0.01)$ & $1.18 (0.00)$ & -& - \\
sh6bench & $1.23 (0.00)$ & $1.06 (0.00)$ & $2.09 (0.00)$ & $1.36 (0.05)$ & 1.01 (0.00)& - \\
sh8bench & $1.02 (0.00)$ & $0.00 (0.00)$ & $1.30 (0.01)$ & $7.30 (0.28)$& 0.97 (0.01) & - \\
xmalloc-test & $3.37 (0.01)$ & $1.01 (0.01)$ & $11.19 (0.01)$ & $13.53 (0.82)$ & 1.03 (0.01)& $7.07 (0.01)$ \\ \midrule
Avg & 1.84 (0.00) & \underline{1.30 (0.01)} & 5.31 (0.01) & 2.16 (0.03) & \textbf{1.03 (0.01)} & 2.08 (0.01)\\
\bottomrule
\end{tabular}

\caption{Normalized average memory overheads (and standard deviations) of \sys on mimalloc-bench (results of * marked tests use built-in measurements). We indicate the best scheme in \textbf{bold} and the second best \underline{underlined} to show how \sys hits the sweet spot in the tradeoff between run time and memory use.}
\label{tab:micro-individual-m}
\end{table*}

\begin{table*}[h]
\small
\centering
\begin{tabular}{l|cccc|cccc}
\toprule
& \multicolumn{4}{c|}{Throughput Overhead (std.)} & \multicolumn{4}{c}{Memory Overhead (std.)} \\ 
& Redis & Nginx & Lighttpd & Average & Redis & Nginx & Lighttpd & Average  \\
\midrule
SEMalloc & 1.06 (0.02) & 0.99 (0.01) & 1.00 (0.01) & \underline{1.02 (0.01)} & 2.22 (0.01) & 2.64 (0.02) & 2.43 (0.02) & 2.42 (0.02) \\
MarkUs & 1.33 (0.04) & 1.12 (0.05) & 1.04 (0.03) & 1.16 (0.04) & 1.77 (0.01) & 303.11 (0.02) & 1.45 (0.02) & 9.20 (0.02) \\
FFMalloc & 1.20 (0.04) & 0.99 (0.02) & 1.00 (0.00) & 1.06 (0.00) & 5.92 (0.02) & 15.48 (0.01) & 11.55 (0.02) & 10.19 (0.02) \\
MineSweeper & 1.21 (0.01) & - & 1.04 (0.00) & 1.12 (0.00) & 2.74 (0.04) & - & 2.53 (0.10) & 2.63 (0.06) \\
TypeAfterType & 1.05 (0.02) & 0.99 (0.00) & 1.00 (0.00) & \textbf{1.01 (0.00)} & 2.91 (0.03) & 1.55  (0.02) & 1.00 (0.02) & \underline{1.65 (0.02)} \\
DangZero &  - & 2.90 (0.01) & - & 2.90 (0.01) & - & 2.37 (0.01) & - & 2.37 (0.01) \\
PUMM & 1.43 (0.01) & 1.01 (0.00) & 1.02 (0.06) & 1.14 (0.00) & 1.03 (0.01) & 1.03 (0.01) & 1.13 (0.01) & \textbf{1.06 (0.01)} \\
\bottomrule
\end{tabular}
\caption{Normalized average throughput and memory overheads (and standard deviations) of \sys on three real-world programs (results of * marked tests use built-in measurements). We indicate the best scheme in \textbf{bold} and the second best \underline{underlined} to show how \sys hits the sweet spot in the tradeoff between run time and memory use.}
\label{tab:realworld-individual}
\end{table*}

\subsection{Maximum Working Set Size (WSS) of
Each Benchmark Test}
\label{app:wss}

\PP{Macro benchmark}
SPEC and PARSEC in~\autoref{tab:macro-wss}.

\PP{Real-world programs}
Three real-world programs
(Redis, Nginx, Lighttpd)
in~\autoref{tab:realworld-wss}.

We note that we do not compare with DangZero 
as most of its memory overheads comes from the page table
management and cannot be reflected using WSS,
and we do not run mimalloc-bench as its tests are time-sensitive and
many tests, such as \cc{glibc-simple}, do not access the allocated memory.

\begin{table*}[h]
\small
\centering
\begin{tabular}{l|ccccc}
\toprule
Test ID & SEMalloc & Markus & FFMalloc & MineSweeper & TypeAfterType \\
\midrule
600 & 0.65 (0.00) & 1.39 (0.74) & 1.59 (0.62) & 1.10 (0.69) & 0.98 (0.03) \\
602 & 1.44 (0.32) & 1.19 (0.11) & 2.69 (0.69) & 1.00 (0.00) & - \\
605 & 1.02 (0.20) & 1.12 (0.18) & 2.96 (1.62) & 1.53 (0.53) & 1.04 (0.03) \\
619 & 0.49 (0.10) & 0.49 (0.10) & 1.72 (0.84) & 1.08 (0.85) & 0.93 (0.06) \\
620 & 1.53 (0.18) & 1.63 (0.29) & 3.37 (1.37) & 1.63 (0.00) & 0.98 (0.06) \\
623 & 1.17 (0.29) & 1.08 (0.14) & 1.92 (0.28) & 1.33 (0.00) & 1.02 (0.03) \\
625 & 1.62 (0.22) & 2.88 (1.82) & 3.00 (0.00) & 1.62 (0.22) & 1.01 (0.00) \\
631 & 3.28 (2.38) & 1.13 (0.23) & 2.15 (0.20) & 3.50 (2.60) & 1.04 (0.00) \\
638 & 2.08 (1.32) & 1.92 (1.61) & 2.42 (0.92) & 1.25 (0.14) & 1.00 (0.00) \\
641 & 0.75 (0.08) & 0.53 (0.00) & 0.62 (0.09) & 0.57 (0.08) & 1.00 (0.03) \\
644 & 1.25 (0.14) & 1.25 (0.14) & 1.92 (0.14) & 2.00 (1.37) & 1.00 (0.00) \\
657 & 3.35 (1.40) & 0.88 (0.14) & 3.03 (1.46) & 0.80 (0.16) & 1.01 (0.00) \\
\midrule
Avg & 1.24 (0.04) & \underline{1.09 (0.05)} & 2.00 (0.18) & 1.21 (0.20) & \textbf{1.00 (0.02)} \\\midrule
bla & 1.40 (0.20) & 1.30 (0.24) & 2.70 (0.24) & 1.40 (0.20) & 1.01 (0.01) \\
bod & 1.13 (0.16) & 1.80 (1.11) & 2.93 (1.18) & 1.27 (0.13) & 1.01 (0.00) \\
can & 1.28 (0.00) & 1.28 (0.00) & 1.87 (0.21) & 1.28 (0.00) & - \\
ded & 1.00 (0.00) & 1.07 (0.13) & 1.60 (0.13) & 1.00 (0.00) & 0.98 (0.01) \\
fer & 1.45 (0.55) & - & 1.51 (0.24) & - & - \\
flu & 1.34 (0.95) & 1.71 (1.16) & 1.62 (0.70) & 0.69 (0.00) & 1.00 (0.00) \\
fre & 1.10 (0.44) & 1.51 (1.02) & 1.76 (0.46) & 1.02 (0.72) & 0.99 (0.01) \\
str & 1.01 (0.15) & 1.07 (0.24) & 1.90 (1.14) & 1.43 (1.22) & 1.03 (0.00) \\
swa & 1.69 (1.53) & 1.69 (1.19) & 3.52 (1.22) & 1.83 (1.13) & - \\
vip & 0.89 (0.09) & 2.77 (0.38) & 1.69 (0.09) & 1.27 (0.90) & - \\
\midrule
Avg & 1.14 (0.16) & 1.39 (0.23) & 1.94 (0.09) & \underline{1.11 (0.09)} & \textbf{1.00 (0.00)}\\
\bottomrule
\end{tabular}
\caption{Normalized maximum WSS (and standard deviations) of SeMalloc on SPEC and PARSEC. We indicate the best scheme in bold and the second best underlined.}
\label{tab:macro-wss}
\end{table*}

\begin{table*}[h]
\small
\centering
\begin{tabular}{l|ccccc}
\toprule
Test ID & SEMalloc & Markus & FFMalloc & MineSweeper & TypeAfterType
\\ \midrule
Redis    & 0.99 (0.32)  & 1.12 (0.04)  & 1.17 (0.08)  & 1.14(0.04) & 1.00 (0.00) \\
Nginx    & 1.02 (0.04)  & 1.08 (0.03)  & 1.03 (0.08)  & -  & 1.00 (0.00) \\
Lighttpd & 0.97 (0.27)  & 0.74 (0.38)  & 1.08 (0.07)  & 1.06 (0.00)  & 1.00 (0.00) \\
Avg      & \underline{0.96 (0.15)}  & \textbf{0.93 (0.18)}  & 1.09 (0.03)  & 1.10(0.02) & 1.00 (0.00) \\ \bottomrule
\end{tabular}
\caption{Normalized maximum WSS (and standard deviations) of SeMalloc on three real-world programs. We indicate the best scheme in bold and the second best underlined.}
\label{tab:realworld-wss}
\end{table*}

\subsection{Recurrent Allocation Statistics
on PARSEC and SPEC Tests}
\label{app:stats}

Allocation details are shown in~\autoref{tab:allocations}.

\begin{table*}[h]
\centering
    \small
    \begin{tabular}{l|rrrrr|rrrrrrr}
    \toprule
        Test 
        & $\Delta$ size 
        & \begin{tabular}[r]{@{}r@{}}\# alloc\\sites\end{tabular} & \begin{tabular}[r]{@{}r@{}}\# CG\\nodes\end{tabular} & \begin{tabular}[r]{@{}r@{}}\# CG\\edges\end{tabular} &
        \begin{tabular}[r]{@{}r@{}}\# CG\\SCCs\end{tabular} &
        \# allocs &
        \begin{tabular}[r]{@{}r@{}}\# rec.\\pools\end{tabular} & \begin{tabular}[r]{@{}r@{}}\# rec.\\allocs\end{tabular} & \begin{tabular}[r]{@{}r@{}}ave. allocs per \\rec. pool\end{tabular} & \begin{tabular}[r]{@{}r@{}} mem. \\leak\end{tabular} & \begin{tabular}[r]{@{}r@{}} \# native \\objs\end{tabular} & \begin{tabular}[r]{@{}r@{}}\# \sema \\objs\end{tabular}  \\
    \midrule
        600* & 8.97\% & 80 & 1,355 & 14,068 & 14 & 47,799,547 & 4,470 & 99.89\% & 10,693 & 0.0\% & 30 & 7943 \\
        602* & 3.06\% & 91 & 2,330 & 23,128 & 12 & 76,739,284 & 1,410 & 99.97\% & 58,990 & 0.0\% & 5 & 1280\\
        605 & 0.06\% & 16 & 9 & 22 & 0 & 1,005,766 & 8 & 100.00\% & 125,719 & 0.0\% & 14 & 14\\
        619 & 1.01\% & 0 & 0 & 0 & 0 & 6 & 0 & 0.00\% & 0 & 0.0\% & 1 & 1\\
        620 & 3.37\% & 1,105 & 1,498 & 110,380 & 5 & 458,738,506 & 698 & 98.60\% & 648,045 & 41.6\% & 567 & 1026\\
        623 & 1.23\% & 30 & 1,387 & 4,273 & 25 & 138,362,610 & 455 & 100.00\% & 304,079 & 1.6\% & 2 & 193\\
        625* & 1.01\% & 66 & 55 & 319 & 0 & 3,245 & 28 & 97.75\% & 113 & 0.6\% & 4 & 87\\
        631 & 0.37\% & 0 & 0 & 0 & 0 & 3 & 0 & 0.00\% & 0 & 0.0\% & 1 & 1\\
        638 & 8.97\% & 10 & 1,249 & 13,271 & 7 & 42,945,318 & 135 & 100.00\% & 613,486 & 0.0\% & 5 & 835\\
        641 & 2.47\% & 90 & 96 & 323 & 0 & 53,759,694 & 648 & 99.87\% & 82,855 & 54.0\% & 35 & 144\\
        644 & 5.17\% & 202 & 83 & 367 & 1 & 1,533,183 & 56 & 100.00\% & 27,378 & 0.0\% & 39 & 57\\
        657* & 1.90\% & 10 & 66 & 221 & 2 & 32 & 5 & 31.25\% & 2 & 0.0\% & 6 & 22 \\ \midrule
        bla & 1.36\% & 0 & 0 & 0 & 0 & 8 & 0 & 0.00\% & 0  & 0.0\% & 1 & 1 \\
        bod & 4.48\% & 55 & 654 & 809 & 0 & 371,834 & 35 & 94.62\% & 10,053 & 34.9\% & 32 & 76 \\
        can & 3.07\% & 3 & 47 & 55 & 0 & 21,141,425 & 12 & 100.00\% & 1,761,785 & 0.0\% & 3 & 5 \\
        ded & 4.98\% & 40 & 41 & 110 & 0 & 1,717,291 & 28 & 100.00\% & 61,328 & 0.0\% & 27 & 27 \\
        fer & 7.78\% & 161 & 125 & 355 & 1 & 521,103 & 104 & 86.5\% & 4,335 & 9.8\% & 76 & 76 \\
        flu & 7.75\% & 12 & 10 & 20 & 0 & 229,910 & 1 & 100.00\% & 229,899 & 0.0\% & 10 & 10 \\
        fre & 10.29\% & 64 & 37 & 150 & 0 & 441 & 6 & 2.72\% & 2 & 0.0\% & 23 & 35 \\
        str & 21.65\% & 22 & 14 & 35 & 0 & 8,835 & 35 & 99.57\% & 251 & 0.0\% & 14 & 15 \\
        swa & 0.84\% & 5 & 12 & 32 & 0 & 48,001,799 & 20 & 100.00\% & 2,400,089 & 0.3\% & 5 & 20 \\
        vip & 24.06\% & 573 & 2,758 & 37,108 & 25 & 2,380,317 & 283 & 99.99\% & 8,410 & 2.5\% & 44 & 320 \\ 
    \bottomrule
    \end{tabular}
    \caption{Number of allocations, iterative allocations, and iterative pools for each SPEC and PARSEC test. 
    We highlight that \sys can efficiently identify \sema{s} and cause negligible 
    memory leakage for most programs.
    Tests with * have more than one input. We only report the input that triggers the most allocations.}
    \label{tab:allocations}
\end{table*}

\end{document}